\documentclass[a4paper,fleqn,usenatbib,useAMS]{mnras}
\usepackage{newtxtext,newtxmath}
\usepackage[T1]{fontenc}
\usepackage{ae,aecompl}
\usepackage{graphicx}

\newcommand       \be           {\begin{equation}}
\newcommand       \ee           {\end{equation}}

\title[The alignment of interstellar dust grains]
{The alignment of interstellar dust grains: thermal flipping and the
Davis-Greenstein mechanism}
\author[J. C. Weingartner, E. Kolasi, \& C. Woods]
{Joseph C. Weingartner$^{1}$\thanks{E-mail: jweinga1@gmu.edu},
Erald Kolasi$^{1}$, and Cameron Woods$^{1}$\\
$^{1}$Department of Physics and Astronomy, George Mason University, 
4400 University Drive, Fairfax, VA 22030, USA}

\date{Accepted XXX. Received YYY; in original form ZZZ}

\pubyear{2021}

\begin{document}
\label{firstpage}
\pagerange{\pageref{firstpage}--\pageref{lastpage}}
\maketitle

\begin{abstract}

Interstellar dust grains are non-spherical and, in some environments, 
partially aligned along the direction of the interstellar magnetic field.
Numerous alignment theories have been proposed, all of which examine the 
grain rotational dynamics. In 1999, Lazarian \& Draine introduced the
important concept of thermal flipping, in which internal relaxation processes 
induce the grain body to flip while its angular momentum remains fixed.
Through detailed numerical simulations, we study the role of thermal
flipping on the grain dynamics during periods of relatively slow rotation,
known as `crossovers', for the special case of a spheroidal grain with a
non-uniform mass distribution. Lazarian \& Draine proposed that rapid flipping
during a crossover would lead to `thermal trapping', in which a systematic
torque, fixed relative to the grain body, would time average to zero, 
delaying spin-up to larger rotational speeds. We find that the time-averaged
systematic torque is not zero during the crossover and that thermal trapping
is not prevalent. As an application, we examine whether the classic 
Davis-Greenstein
alignment mechanism is viable, for grains residing in the cold neutral medium
and lacking superparamagnetic inclusions. We find that Davis-Greenstein 
alignment is not hindered by thermal trapping, but argue that it is,
nevertheless, too inefficient to yield the alignment of large grains 
responsible for optical and infrared starlight polarization. Davis-Greenstein
alignment of small grains could potentially contribute to the observed
ultraviolet polarization. The theoretical and computational tools developed
here can also be applied to analyses of alignment via radiative torques and 
rotational disruption of grains.

\end{abstract}


\section{Introduction} \label{sec:intro}

In the classic Davis-Greenstein (`D-G') theory \citep{DG51}, paramagnetic
dissipation in rotating grains drives the grains 
into alignment with the interstellar
magnetic field. However, disalignment due to random collisions with gas-phase
particles renders this mechanism ineffective if the grain rotation is
excited by those same collisions. 

\citet{Purcell79} noted that grains are subject to systematic
torques, fixed relative to the grain body, that can potentially drive
them to suprathermal rotation. That is, the grain's angular speed could 
exceed, by a factor of several or more, the thermal rotation rate 
$\omega_T$ that results when random collisions with gas particles excite
the rotation. Suprathermally rotating grains can avoid being disaligned by
these same random collisions. 

\citet{Purcell79} also noted that internal mechanisms can dissipate 
rotational kinetic energy into heat within the grain, driving it to rotate
around its principal axis of greatest moment of inertia, henceforth denoted
$\bmath{\hat{a}}_1$. \citet{Purcell79} described a previously unexamined
mechanism,
`Barnett dissipation', in which a paramagnetic grain attempts to magnetize
along the direction of the Barnett-equivalent field 
$\mathbfit{B}_{\mathrm{BE}} = \bmath{\omega}/\gamma_g$. Here $\bmath{\omega}$
is the grain's angular velocity vector, which rapidly varies as observed in
a reference frame fixed to the grain when the grain does not rotate about
a principal axis, and $\gamma_g$ is the gyromagnetic ratio of the microscopic 
spins that give rise to the grain's paramagnetism. \citet{Purcell79} 
considered dissipation associated with electron paramagnetism. 
\citet{LD99b} described the related phenomenon of `nuclear relaxation',
associated with nuclei such as H that are likely incorporated within grains,
and found that it can be much more efficient than Barnett dissipation for 
thermally rotating grains of a wide range of sizes.

The dominant systematic torque considered by \citet{Purcell79} is due to 
the recoil from H$_2$ molecules that form on the grain surface and  
are subsequently ejected into the gas, with some of the released binding
energy converted to translational kinetic energy. If the molecules form only at
certain special sites on the surface, then the net recoil torque is 
non-zero. In this work, we neglect gas-grain drift and photodesorption of 
adatoms; with these assumptions, the net recoil torque is fixed relative to 
the grain body. The special
sites are not permanent; new sites form and old sites disappear as the grain
undergoes resurfacing, e.g.~due to the accretion of atoms from the gas. As
a result, the component $\Gamma_1$ of the mean systematic torque along 
$\bmath{\hat{a}}_1$ can sometimes change sign. For a suprathermally rotating 
grain, in which Barnett dissipation is highly efficient, $\bmath{\omega}$ is 
either nearly parallel to or nearly anti-parallel to $\bmath{\hat{a}}_1$. 
Thus, when $\Gamma_1$ changes sign, the grain enters a period of spin-down,
and may ultimately spend some time rotating thermally. During these episodes,
known as `crossovers', the grain is again susceptible to disalignment via
random collisions with gas particles. 
In the first study of crossovers, \citet{SM79} concluded that a grain would 
become disaligned after passing through a small number of crossovers. 

In the inverse process of Barnett dissipation, 
thermal fluctuations prevent $\bmath{\omega}$
from lying exactly along $\bmath{\hat{a}}_1$ during periods of suprathermal
rotation. Surprisingly, \citet{LD97} found that this limits the extent of 
disalignment during crossovers. However, \citet{LD99a} concluded that Barnett
fluctuations are so strong during periods of thermal rotation that a grain
can flip. That is, the sign of $\mathbfit{J} \bmath{\cdot} \bmath{\hat{a}}_1$
can change as a result of thermal fluctuations; $\mathbfit{J}$ is the grain's
angular momentum. \citet{LD99a} called this process `thermal flipping'.

When the grain flips, so does the direction of the systematic torque relative
to the angular momentum vector $\mathbfit{J}$. Denoting the interval
between consecutive flips as an `f-step', the time-averaged sytematic torque 
will equal zero if the mean f-step duration is the same for the case that
$\mathbfit{J} \bmath{\cdot} \bmath{\hat{a}}_1 < 0$ as for the case that
$\mathbfit{J} \bmath{\cdot} \bmath{\hat{a}}_1 > 0$. (In making this statement,
we neglect any grain resurfacing during a crossover, as well as the
dependence of the systematic torque on the grain kinetic energy at fixed
angular momentum, and assume that the f-step duration is much less than the
crossover duration.) As a result, the time required for the grain to emerge
from a crossover may be much longer than for the 
case where thermal flipping does not occur
and the effect of the systematic torque compounds uniformly over time. 
\citet{LD99a} called this phenomenon `thermal trapping'.
Analyzing nuclear relaxation, \citet{LD99b} concluded that all grains that
contribute to starlight polarization would likely undergo thermal trapping,
arguing against Davis-Greenstein alignment aided by suprathermal spin-up
due to torques fixed relative to the grain body. 

For simplicity, the above studies focused on oblate grains exhibiting dynamic
symmetry. That is, $I_1 > I_2 = I_3$, where $I_i$ is the moment of inertia
associated with principal axis $\bmath{\hat{a}}_i$. \citet{W09} showed that
thermal flipping does not actually occur for grains with dynamic symmetry.
\citet{W09} noted that external processes (e.g.~collisions with gas atoms
and the ejection of H$_2$ molecules from the grain surface) might induce
grain flipping; \citet{HL09} examined this possibility quantitatively.

\citet{KW17} showed that thermal flipping does
occur for grains that lack dynamic symmetry. They only provided quantitative
results for the relaxation rate for Barnett relaxation. In Section 
\ref{sec:nuclear} of this work, we extend their analysis to also treat nuclear 
relaxation.

Furthermore, we challenge the assumption noted above, that the mean f-step
duration is the same for the case that
$\mathbfit{J} \bmath{\cdot} \bmath{\hat{a}}_1 < 0$ as for the case that
$\mathbfit{J} \bmath{\cdot} \bmath{\hat{a}}_1 > 0$. Because of the systematic
torque, the magnitude $J$ of the angular momentum is more likely to increase
when $\mathbfit{J} \bmath{\cdot} \Gamma_1 \bmath{\hat{a}}_1 > 0$ and is more
likely to decrease when
$\mathbfit{J} \bmath{\cdot} \Gamma_1 \bmath{\hat{a}}_1 < 0$. We will refer to
an f-step with $\mathbfit{J} \bmath{\cdot} \Gamma_1 \bmath{\hat{a}}_1 > 0$
as an `up-step' and an f-step with
$\mathbfit{J} \bmath{\cdot} \Gamma_1 \bmath{\hat{a}}_1 < 0$ as a `down-step'.
(Note that $J$ can decrease during an up-step or increase during a down-step
since gas-atom collisions and H$_2$-formation events occur stochastically.)
In addition, the flipping probability per unit time decreases as $J$ increases.
Thus, the mean f-step duration is longer for f-steps in which $J$ increases
than for f-steps in which $J$ decreases. We conclude that the mean duration is
longer for up-steps than for down-steps. In other words, there is
a bias for f-steps to have longer duration when the systematic torque acts
to spin the grain up than when it acts to spin the grain down. Consequently,
the systematic torque does not average to zero, potentially circumventing
thermal trapping. The main purpose of this work is to examine this
possibility, employing detailed numerical simulations. 

For simplicity, we consider an oblate spheroidal grain with an inhomogeneous 
mass distribution. Thus, the grain exhibits geometric symmetry but not dynamic
symmetry. We assume that the mass is distributed such that the grain's center 
of mass remains at its geometric center and the principal axis of largest
moment of inertia, $\bmath{\hat{a}}_1$, lies along the symmetry axis, but
$I_2 \ne I_3$. 

The equations and quantities used in the simulations are derived in Sections
\ref{sec:spheroid-properties}--\ref{sec:external-processes}. Section 
\ref{sec:simulations} describes the fundamental elements of the simulations.
Section \ref{sec:code-test}
presents a test of the simulation code. Two simplified versions of the code
are used to examine the case where the grain is in thermal equilibrium, for
which analytic results are available for comparison. The codes successfully
reproduce the expected results. Section \ref{sec:H2-formation}
describes our treatment of H$_2$ formation (and subsequent ejection) at special
surface sites. Sections \ref{sec:bias}--\ref{sec:D-G-alignment} contain our
main simulation results, focusing on grain dynamics in the cold neutral medium
of the diffuse ISM. Section \ref{sec:bias} presents simulation results for one
specific case, demonstrating that the mean up-step duration exceeds the mean
down-step duration in that case. Section \ref{sec:crossovers} presents
results from large suites of simulations examining crossovers, specifically
their duration and the associated grain disorientation. We find that thermal
trapping is not prevalent. The simulations in Section \ref{sec:D-G-alignment}
examine the efficiency of Davis-Greenstein alignment in the cold neutral 
medium. We find that thermal trapping does not inhibit suprathermal spin-up and
D-G alignment, but the alignment time-scale (for grains lacking 
superparamagnetic inclusions) is uncomfortably long. Conclusions and future
work are summarized in Section \ref{sec:conclusions}.

\section{Nuclear relaxation in non-symmetric grains} \label{sec:nuclear}

Consider a non-symmetric grain with $I_1 > I_2 > I_3$ and denote
$r_2 = I_1/I_2$ and $r_3 = I_1/I_3$. It is convenient to define a dimensionless
measure of the grain's rotational kinetic energy $E$ (when $J$ is constant),
\be
\label{eq:q}
q = \frac{2 I_1 E}{J^2} .
\ee
The kinetic energy is minimized (maximized) for rotation about
$\bmath{\hat{a}}_1$ ($\bmath{\hat{a}}_3$); thus, $1 \le q \le r_3$.
In the absence of external processes, the evolution of $q$ due to internal
relaxation can be described using the Langevin equation,
\be
\label{eq:internal-Langevin-eq}
\mathrm{d}q = A(q) \, \mathrm{d}t + \sqrt{D(q)} \, \mathrm{d}w_{\mathrm{int}} ,
\ee
where $\mathrm{d}t$ is a time step, $\mathrm{d}w_{\mathrm{int}}$ is a Gaussian
random variable with variance $\mathrm{d}t$, the drift coefficient $A(q)$ is
proportional to the average energy dissipation rate, and $D(q)$ is the
diffusion coefficient.

\citet{KW17} examined Barnett relaxation in non-symmetric
grains, deriving expressions for the drift and diffusion coefficients in the
low-frequency limit, i.e.~when the angular frequency of the grain's rotation
is much less than $T_2^{-1}$, where $T_2$ is the spin-spin relaxation time.
For Barnett relaxation, associated with electrons in, e.g., Fe atoms, this
approximation is excellent. However, in the case of nuclear relaxation, it
can fail for thermally rotating grains. Nuclear relaxation dominates
Barnett relaxation for most thermally rotating grains. Thus, in this
section, we relax the low-frequency limit.
Readers who are not familiar with \citet{KW17} may want to proceed
directly to the results in the final two paragraphs of this section.

The only term in the \citet{KW17} expression for the drift coefficient $A(q)$ 
that changes when the low-frequency limit is relaxed is the following function,
which is equivalent to the expression on the second line in their equation
(49): 
\begin{multline}
\Theta(q, \Psi) = \left( T_2^{\prime} \right)^{-2} \int_0^{4 K(k^{\pm 2})} \mathrm{d}u
\left\{
c_1(q) \mathcal{F}_{\pm}\left(u, k^{\pm 2}\right) \times \right. \\
\left[ c_1(q) \mathcal{F}_{\pm}
\left(u, k^{\pm 2}\right) - M_1^{\prime}(q, u) \right] 
+ c_2(q) \mathrm{sn} \left(u, k^{\pm 2}\right) \times \\
\left[ c_2(q) \mathrm{sn} \left(u, k^{\pm 2}\right) - M_2^{\prime}(q, u) \right]
+ c_3(q) \mathcal{F}_{\mp}\left(u, k^{\pm 2}\right) \times \\
\left. \left[ c_3(q) \mathcal{F}_{\mp}
\left(u, k^{\pm 2}\right) - M_3^{\prime}(q, u) \right] \right\} 
\label{eq:define-Theta}
\end{multline}
where the $+$ ($-$) sign in $k^{\pm 2}$ and $\mathcal{F}_{\pm}$
is for $1 < q < r_2$ ($r_2 < q < r_3$);
\be
T_2^{\prime} = \Psi \times \begin{cases}
\left[ (r_2 - 1) (r_3 - q) \right]^{1/2} & 1 < q < r_2 \\
\left[ (r_3 - r_2) (q-1) \right]^{1/2} & r_2 < q < r_3
\end{cases} ;
\ee
\be
\Psi = \frac{J T_2}{I_1} ;
\ee
$K(k^{\pm 2})$ is the complete elliptic integral of the first kind,
\be
k^2 = \frac{(r_3 - r_2) (q-1)}{(r_2 -1) (r_3 -q)} ;
\ee
$\mathcal{F}_+(u, k^{\pm 2}) = \mathrm{dn}(u, k^{\pm 2})$;
$\mathcal{F}_-(u, k^{\pm 2}) = \mathrm{cn}(u, k^{\pm 2})$;
$\mathrm{sn}(u, k^{\pm 2})$, $\mathrm{cn}(u, k^{\pm 2})$, and
$\mathrm{dn}(u, k^{\pm 2})$, are the Jacobi elliptic functions; 
\be
c_1(q) = \begin{cases}
[(r_2-1) (r_3-1)]^{-1/2} & , 1 < q < r_2 \\
(r_3-q)^{1/2} [(r_3-1) (r_3-r_2) (q-1)]^{-1/2} & , r_2 < q < r_3
\end{cases} ;
\ee
\be
c_2(q) = \begin{cases}
- r_2 (q-1)^{1/2} (r_2-1)^{-1} (r_3-q)^{-1/2} & , 1 < q < r_2 \\
- r_2 (r_3-q)^{1/2} (r_3-r_2)^{-1} (q-1)^{-1/2} & , r_2 < q < r_3
\end{cases} ;
\ee
\be
c_3(q) = r_3 \left( \frac{q-1}{r_3 -q} \right)^{1/2} c_1(q) ;
\ee
and $M_i^{\prime}(q, u)$ are the steady-state solutions of the following
differential equations:
\begin{multline}
\frac{\mathrm{d}
M_1^{\prime}}{\mathrm{d}u} = c_3(q) M_2^{\prime}(q,u) \mathcal{F}_{\mp}(u, k^{\pm 2})
- c_2(q) M_3^{\prime}(q,u) \mathrm{sn}(u, k^{\pm 2}) \\
+ \left(T_2^{\prime}\right)^{-1}
\left[ c_1(q) \mathcal{F}_{\pm}(u, k^{\pm 2}) - M_1^{\prime}(u) \right] ,
\label{eq:modified-Bloch1}
\end{multline}
\begin{multline}
\frac{\mathrm{d}
M_2^{\prime}}{\mathrm{d}u} = c_1(q) M_3^{\prime}(q,u) \mathcal{F}_{\pm}(u, k^{\pm 2})
- c_3(q) M_1^{\prime}(q,u) \mathcal{F}_{\mp}(u, k^{\pm 2}) \\
+ \left(T_2^{\prime}
\right)^{-1} \left[ c_2(q) \mathrm{sn}(u, k^{\pm 2}) - M_2^{\prime}(u) \right] ,
\end{multline}
\begin{multline}
\frac{\mathrm{d}M_3^{\prime}}{\mathrm{d}u} 
= c_2(q) M_1^{\prime}(q,u) \mathrm{sn}(u, k^{\pm 2})
- c_1(q) M_2^{\prime}(q,u) \mathcal{F}_{\pm}(u, k^{\pm 2}) \\
+ \left(T_2^{\prime}
\right)^{-1} \left[ c_3(q) \mathcal{F}_{\mp}(u, k^{\pm 2}) - M_3^{\prime}(u) \right]
.
\label{eq:modified-Bloch3}
\end{multline}
We adopt the same conventions for the Jacobi elliptic functions as 
\citet{WD03}. 

From \citet{KW17}, in the low-frequency limit
\be
\Theta(q, \Psi \ll 1) =
\frac{4 \left\{ z_1 [E(k^2) + (k^2-1)
K(k^2)] + k^2 z_2 E(k^2) \right\} (q-1)}{3 k^2 (r_2-1)^2 (r_3-1) (r_3-q)}
\ee
when $1 < q < r_2$ and
\begin{multline}
\Theta(q, \Psi \ll 1) =
4 \left\{ z_2 [E(k^{-2}) + (k^{-2}-1) K(k^{-2})] + k^{-2} z_1 E(k^{-2}) 
\right\} \\
\times \frac{(r_3 - q)}{3 k^{-2} (r_3 - r_2)^2 (r_3-1) (q-1)}
\end{multline}
when $r_2 < q < r_3$; 
$E(k^2)$ is the complete elliptic integral of the second kind,
\be
z_1 = 2 (r_3-r_2) - r_3^2 (r_2-1) + r_2^2 (r_3-1) ,
\ee 
and
\be
z_2 = - (r_3-r_2) +2 r_3^2 (r_2-1) + r_2^2 (r_3-1) .
\ee

For the specific case that $r_2 = 1.3$ and $r_3 = 1.5$, we find the 
steady-state solution of equations 
(\ref{eq:modified-Bloch1})--(\ref{eq:modified-Bloch3}) and perform the
integration in equation (\ref{eq:define-Theta}) numerically. We
find that $\Theta(q, \Psi)$ is very close to $\Theta(q, \Psi \ll 1)$
when $\Psi \la 0.1$ and $\Theta(q, \Psi)/\Theta(q, \Psi \ll 1)$ drops
to $\approx 6 \times 10^{-3}$ when $\Psi = 10$. For a given value of 
$\Psi$, $\Theta(q, \Psi)/\Theta(q, \Psi \ll 1)$ varies by less than 
30 per cent as $q$ ranges from 1 to $r_3$. 

\citet{KW17} expressed the drift coefficient $A(q)$ in the low-frequency 
limit in the form $A(q) = - \tau^{-1}_{\mathrm{int}} A_1(q)$, where $A_1(q)$ is
a dimensionless function of $q$, given in equations (55) and (61) in 
\citet{KW17}, and $\tau_{\mathrm{int}}$ is the internal relaxation 
time-scale.
Given the gross uncertainties in the
theoretical modeling of Barnett and nuclear relaxation, we do not adjust the
functional form of $A_1(q)$ from its low-frequency form.
Rather, we simply adjust the relaxation time-scale according to
\be
\label{eq:tau-int-all-freq}
\tau_{\mathrm{int}}(\Psi) = \tau_{\mathrm{int}}(\Psi \ll 1) \ \left[
\frac{\Theta(q, \Psi \ll 1)}{\Theta(q, \Psi)} \right]_{\mathrm{av}} .
\ee
In the final term in equation (\ref{eq:tau-int-all-freq}), the ratio
$\Theta(q, \Psi \ll 1)/\Theta(q, \Psi)$ is averaged over $q$ for a fixed value
of $\Psi$. To within 0.5 per cent,
\be
\label{eq:tau-nuc-factor}
\left[ \frac{\Theta(q, \Psi \ll 1)}{\Theta(q, \Psi)} \right]_{\mathrm{av}} = 
\left( 1 + 1.67 \, \Psi^{1.96} \right)^{1.02} .
\ee
Note that this fit is specifically for grains with $r_2 = 1.3$ and 
$r_3 = 1.5$. 

From equation (56) in \citet{KW17}, 
\be
\label{eq:tau-int-low-freq}
\tau_{\mathrm{int}}(\Psi \ll 1) = \frac{\gamma_g^2 I_1^3}{2 \chi_0 V T_2 J^2} .
\ee
As in \citet{WD03}, we take 
$\gamma_g = -1.76 \times 10^7 \, \mathrm{s}^{-1} \, \mathrm{G}^{-1}$ and
$\chi_0 T_2 = 10^{-13} (15 \, \mathrm{K}/T_d) \, \mathrm{s}$ for Barnett 
relaxation ($T_d$ is the dust temperature) and 
$\gamma_g = 1.3 \times 10^4 \, \mathrm{s}^{-1} \, \mathrm{G}^{-1}$,
$\chi_0 = 4 \times 10^{-11} (15 \, \mathrm{K}/T_d)$, and 
$T_2 = 10^{-4} \, \mathrm{s}$ for nuclear relaxation. For Barnett relaxation,
we assume that the low-frequency limit always applies.

\section{Spheroid Properties} \label{sec:spheroid-properties}

Consider a spheroid characterized by the radius $a_{\mathrm{eff}}$
of a sphere with equal volume, the ratio $\delta$ of the semilength $a$ along
the (geometric) symmetry axis to the semilength $b$ along a perpendicular 
axis, and the
average mass density $\bar{\rho}$. (Recall that we take the density to vary 
throughout the grain, so that dynamic symmetry is violated.) 
When $\delta > 1$ the spheroid is prolate and when
$\delta < 1$ the spheroid is oblate. In either case, 
$a = a_{\mathrm{eff}} \, \delta^{2/3}$.  Denote the principal axes 
$\bmath{\hat{a}}_1$, $\bmath{\hat{a}}_2$, and $\bmath{\hat{a}}_3$, with 
corresponding moments of inertia 
\be
\label{eq:I_i}
I_i = \frac{8}{15} \, \upi \bar{\rho} a_{\mathrm{eff}}^5 \, \alpha_i  , 
\ee
and take the symmetry axis to lie along $\bmath{\hat{a}}_1$. We take 
$\alpha_1 = \delta^{-2/3}$, its value for a uniform spheroid, and assign
smaller, but unequal, values to both $\alpha_2$ and $\alpha_3$. For a 
uniform spheroid, these would be given by 
$\alpha_2 = \alpha_3 = (\alpha_1 + \alpha_1^{-2})/2$. Here, we consider an
oblate grain with $\delta = 0.5$ and take  
$r_2 = \alpha_1/\alpha_2 = 1.3$ and $r_3 = \alpha_1/\alpha_3 = 1.5$.
(One simple, albeit unrealistic, mass distribution reproducing these values
for $\alpha_1$, $r_2$, and $r_3$ consists of a uniform mass distribution 
throughout the spheroid plus three point particles located at the surface of
the grain along the $a_1$-, $a_2$-, and $a_3$-axes, with mass fractions,
i.e.~the mass of the point mass divided by the entire mass of the grain, of
0.2229, 0.0538, and 0.0949, respectively.)

It will be convenient to also denote 
$(\bmath{\hat{a}}_2, \bmath{\hat{a}}_3, \bmath{\hat{a}}_1)$ by
$(\bmath{\hat{x}}, \bmath{\hat{y}}, \bmath{\hat{z}})$. 
In order to evaluate mean torques and diffusion coefficients associated 
with gas-atom collisions and H$_2$ formation, it is most convenient to 
adopt oblate spheroidal coordinates $(\eta, \phi^{\prime})$.  The 
transformation to grain-body Cartesian coordinates is
\be
\label{eq:x-oblate-spheroidal}
x = a_{\mathrm{eff}} \, \delta^{-1/3} \cos \eta \cos \phi^{\prime} ,
\ee
\be
y = a_{\mathrm{eff}} \, \delta^{-1/3} \cos \eta \sin \phi^{\prime} ,
\ee
\be
z = a_{\mathrm{eff}} \, \delta^{2/3} \sin \eta ,
\label{eq:z-oblate-spheroidal}
\ee
and $-\upi/2 \le \eta \le \upi/2$, $0 \le \phi^{\prime} < 2 \upi$.  
The surface area element is 
\be
\label{eq:surf-area-element}
\mathrm{d}S = 
a_{\mathrm{eff}}^2 \, \delta^{-2/3} \left[ \delta^2 + \left(1-\delta^2 \right) 
\sin^2 \eta \right]^{1/2} \cos \eta \, \mathrm{d}\eta \, \mathrm{d}\phi^{\prime}
\ee
and the outward-pointing unit normal is
\begin{multline}
\bmath{\hat{N}} = \left[ \delta^2 + \left(1-\delta^2 \right) \sin^2 \eta
\right]^{-1/2} \times \\
\left[
\delta \cos \eta \, \left(\bmath{\hat{x}} \cos \phi^{\prime} + \bmath{\hat{y}}
\sin \phi^{\prime} \right) + \bmath{\hat{z}} \sin \eta \right] .
\label{eq:N-hat}
\end{multline}
Along with $\bmath{\hat{N}}$, the following two vectors form an orthonormal
basis:
\be
\bmath{\hat{\phi}^{\prime}} = - \bmath{\hat{x}} \sin \phi^{\prime} +
\bmath{\hat{y}} \cos \phi^{\prime} ,
\ee
\begin{multline}
\bmath{\hat{t}} = \bmath{\hat{\phi}^{\prime}} \bmath{\times} \bmath{\hat{N}} 
= \left[ \delta^2 + \left(1-\delta^2 \right) \sin^2 \eta \right]^{-1/2} \times \\
\left[ \sin \eta \, \left(\bmath{\hat{x}} \cos
\phi^{\prime} + \bmath{\hat{y}} \sin \phi^{\prime} \right) - \bmath{\hat{z}} \delta
\cos \eta \right] .
\label{eq:t-hat}
\end{multline}

For later use, we define the following integrals over the coordinate $\eta$:
\be
\label{eq:app-i1}
\mathcal{I}_1(\delta) =
\int_{-\upi/2}^{\upi/2} \mathrm{d}\eta \, \cos \eta \, [A(\delta, \eta)]^{-1}
\sin^2 \eta = \frac{2 - \delta^2 - \delta^4 g(\delta)}{4 (1 - \delta^2)} ,
\ee
\be
\label{eq:app-i2}
\mathcal{I}_2(\delta) =
\int_{-\upi/2}^{\upi/2} \mathrm{d}\eta \, \cos \eta \, [A(\delta, \eta)]^{-1} =
1 + \delta^2 g(\delta) ,
\ee
\be
\label{eq:app-i3}
\mathcal{I}_3(\delta) = \int_{-\upi/2}^{\upi/2} \mathrm{d}\eta \, \cos \eta \,
A(\delta, \eta) = 2 g(\delta) ,
\ee
\begin{multline}
\mathcal{I}_4(\delta) = \int_{-\upi/2}^{\upi/2} \mathrm{d}\eta \, \cos \eta \,
A(\delta, \eta) \sin^2 \eta \cos^2 \eta \\
= \frac{2 + \delta^2 - \delta^2 (4 - \delta^2) g(\delta)}{4 (1 - \delta^2)^2} ,
\label{eq:app-i4}
\end{multline}
where
\be
\label{eq:A-delta-eta}
A(\delta, \eta) = \left[ \delta^2 + \left( 1 - \delta^2 \right) \sin^2 \eta
\right]^{-1/2} 
\ee
and 
\be
g(\delta) = \frac{1}{2} \left( 1 - \delta^2 \right)^{-1/2} \ln \left[
\frac{1 + (1 - \delta^2)^{-1/2}}{-1 + (1 - \delta^2)^{-1/2}} \right] .
\ee

\section{Coordinate systems}

\label{sec:coord-systems}

We already introduced grain-body coordinates $(x,y,z)$, fixed with respect
to the grain, in Section \ref{sec:spheroid-properties}. Now consider an inertial
coordinate system, which we call `alignment coordinates'
$(x_B, y_B, z_B)$, with its origin also at the center of the spheroidal grain. 

The orientation of the grain in space depends on its angular momentum 
$\mathbfit{J}$ and rotational kinetic energy $E$.  We denote the spherical 
coordinates of $\mathbfit{J}$ in alignment coordinates by 
$(J, \xi, \phi_B)$.  We define `angular-momentum coordinates'
$(x_J, y_J, z_J)$ by
\be
\label{eq:x-J}
\bmath{\hat{x}}_J = \bmath{\hat{\xi}} = \bmath{\hat{x}}_B \cos \xi \cos
\phi_B + \bmath{\hat{y}}_B \cos \xi \sin \phi_B - \bmath{\hat{z}}_B \sin \xi ,
\ee
\be
\bmath{\hat{y}}_J = \bmath{\hat{\phi}}_B = - \bmath{\hat{x}}_B \sin \phi_B
+ \bmath{\hat{y}}_B \cos \phi_B ,
\ee
\be
\bmath{\hat{z}}_J = \bmath{\hat{J}} = \bmath{\hat{x}}_B \sin \xi \cos
\phi_B + \bmath{\hat{y}}_B \sin \xi \sin \phi_B + \bmath{\hat{z}}_B \cos \xi .
\label{eq:z-J}
\ee

The orientation of the grain body in angular-momentum coordinates can be
expressed using Eulerian angles $(\alpha, \gamma, \zeta)$.  
We adopt the same prescription for the Eulerian angles as in section 2.5.3 in 
\citet{WD03}:  
Start with the grain axes
$(\bmath{\hat{a}}_2, \bmath{\hat{a}}_3, \bmath{\hat{a}}_1)$ aligned 
with $(\bmath{\hat{x}}_J, \bmath{\hat{y}}_J, \bmath{\hat{z}}_J)$.
Then apply the following operations
to the grain:  (1) rotate through angle $\zeta$ about 
$\bmath{\hat{a}}_1 = \bmath{\hat{z}}_J$, (2) rotate through angle
$\gamma$ about $\bmath{\hat{a}}_2$, (3) rotate through angle $\alpha$ about
$\bmath{\hat{a}}_1$.  Thus, the transformation between grain-body and 
angular-momentum coordinates is
\begin{multline}
\bmath{\hat{x}} = \bmath{\hat{a}}_2 = \bmath{\hat{x}}_J (\cos \alpha \cos
\zeta - \sin \alpha \sin \zeta \cos \gamma) \\
+ \bmath{\hat{y}}_J (\cos \alpha
\sin \zeta + \sin \alpha \cos \zeta \cos \gamma) + 
\bmath{\hat{z}}_J \sin \alpha \sin \gamma ,
\label{eq:x-hat}
\end{multline}
\begin{multline}
\bmath{\hat{y}} = \bmath{\hat{a}}_3 = - \bmath{\hat{x}}_J (\sin \alpha \cos
\zeta + \cos \alpha \sin \zeta \cos \gamma) \\
+ \bmath{\hat{y}}_J (\cos \alpha
\cos \zeta \cos \gamma - \sin \alpha \sin \zeta) + \bmath{\hat{z}}_J \cos
\alpha \sin \gamma ,
\label{eq:y-hat}
\end{multline}
\be
\bmath{\hat{z}} = \bmath{\hat{a}}_1 = \bmath{\hat{x}}_J \sin \zeta \sin \gamma
- \bmath{\hat{y}}_J \cos \zeta \sin \gamma + \bmath{\hat{z}}_J \cos \gamma .
\label{eq:z-hat}
\ee

\section{Grain rotation}

\label{sec:grain-rotation}

A description of the free rotation of a non-symmetric grain, for a given
$J$ and $q$, and its flipping dynamics can be found in section 2.5 of
\citet{WD03}. As described there, the components of the grain's
angular velocity $\bmath{\omega}$ along the principal axes involve the Jacobi
elliptic functions. For later convenience, we reproduce the expressions for
$\omega_i$ from \citet{KW17} here. 
When
$1 < q < r_2$,
\be
\label{eq:omega1-low-q}
\omega_1 = \pm \frac{J}{I_1} \left( \frac{r_3-q}{r_3-1} \right)^{1/2} 
\mathrm{dn}(\omega_{\mathrm{rot}} t, k^2) ,
\ee
\be
\label{eq:omega2-low-q}
\omega_2 = - \frac{J}{I_1} r_2 \left( \frac{q-1}{r_2-1} \right)^{1/2} 
\mathrm{sn}(\omega_{\mathrm{rot}} t, k^2) ,
\ee
\be
\omega_3 = \pm \frac{J}{I_1} r_3 \left( \frac{q-1}{r_3-1} \right)^{1/2} 
\mathrm{cn}(\omega_{\mathrm{rot}} t, k^2) ,
\label{eq:omega3-low-q}
\ee
where
\be
\label{eq:k2}
k^2 = \frac{(r_3-r_2) (q-1)}{(r_2-1) (r_3-q)}
\ee
and
\be
\omega_{\mathrm{rot}} = \frac{J}{I_1} \left[ (r_2-1) (r_3-q) \right]^{1/2} .
\ee
The grain is in the positive flip state with respect to $\bmath{\hat{a}}_1$
(i.e.~$\mathbfit{J} \bmath{\cdot} \bmath{\hat{a}}_1 > 0$) when the plus sign
is chosen
in both equations (\ref{eq:omega1-low-q}) and (\ref{eq:omega3-low-q}). It is
in the negative flip state with respect to $\bmath{\hat{a}}_1$ when the minus
sign is chosen in both of those cases. When $r_2 < q < r_3$,
\be
\label{eq:omega1-high-q}
\omega_1 = \pm \frac{J}{I_1} \left( \frac{r_3-q}{r_3-1} \right)^{1/2} 
\mathrm{cn}(\omega_{\mathrm{rot}} t, k^{-2}) ,
\ee
\be
\label{eq:omega2-high-q}
\omega_2 = - \frac{J}{I_1} r_2 \left( \frac{r_3 - q}{r_3-r_2} \right)^{1/2} 
\mathrm{sn}(\omega_{\mathrm{rot}} t, k^{-2}) ,
\ee
\be
\omega_3 = \pm \frac{J}{I_1} r_3 \left( \frac{q-1}{r_3-1} \right)^{1/2} 
\mathrm{dn}(\omega_{\mathrm{rot}} t, k^{-2}) ,
\label{eq:omega3-high-q}
\ee
with $k^2$ as defined in equation (\ref{eq:k2}) and 
\be
\omega_{\mathrm{rot}} = \frac{J}{I_1} \left[ (r_3-r_2) (q-1) \right]^{1/2} .
\ee
The grain is in the positive flip state with respect to $\bmath{\hat{a}}_3$
(i.e.~$\mathbfit{J} \bmath{\cdot} \bmath{\hat{a}}_3 > 0$) when the plus sign
is chosen
in both equations (\ref{eq:omega1-high-q}) and (\ref{eq:omega3-high-q}). It is
in the negative flip state with respect to $\bmath{\hat{a}}_3$ when the minus
sign is chosen in both of those cases.

The components $\omega_i$ can also be expressed in terms of the Eulerian angles:
\be
\label{eq:omega_1}
\omega_1 = \frac{J}{I_1} \cos \gamma ,
\ee
\be
\label{eq:omega_2}
\omega_2 = \frac{J}{I_2} \sin \gamma \sin \alpha ,
\ee
\be
\label{eq:omega_3}
\omega_3 = \frac{J}{I_3} \sin \gamma \cos \alpha .
\ee

As seen in equations (\ref{eq:omega1-low-q})--(\ref{eq:omega3-low-q}) and
(\ref{eq:omega1-high-q})--(\ref{eq:omega3-high-q}), the angular velocity in
grain-body coordinates is periodic in variable $\nu = \omega_{\mathrm{rot}} t$, 
with period $4 K(k^{\pm 2})$, where the + (-) sign is for the case that
$1 < q < r_2$ ($r_2 < q < r_3$). Thus, from equations
(\ref{eq:omega_1})--(\ref{eq:omega_3}), the Eulerian angles $\alpha$ and
$\gamma$ are likewise periodic. The Eulerian angle $\zeta$ can be expressed
as the sum of two periodic functions, one with the same period as for
$\alpha$ and $\gamma$ and the other with an incommensurate period. When
evaluating drift and diffusion coefficients associated with external
processes, we will average over the grain rotation, since the rotation
time-scale is orders of magnitude smaller than all other relevant
time-scales. Denoting the average of a function $F$ over grain
rotation by $\langle F \rangle$,
\be
\label{eq:avg-grain-rot}
\langle F \rangle = \left[ 8 \upi K \left( k^{\pm 2} \right) \right]^{-1}
\int_0^{4 K(k^{\pm 2})} \mathrm{d}\nu \int_0^{2 \upi} \mathrm{d}\zeta \, F
\left[ \alpha(\nu), \gamma(\nu), \zeta \right] .
\ee

For a grain with dynamic symmetry, $\gamma$ is constant when $J$ and $E$ are
fixed. As seen above, $\gamma$ varies periodically for a grain that lacks
dynamic symmetry. From equations (\ref{eq:omega_1}),
(\ref{eq:omega1-low-q}), and (\ref{eq:omega1-high-q}), the average value of 
$\cos^2 \gamma$ is given by
\be
\label{eq:avg-cos2-gamma}
\langle \cos^2 \gamma \rangle = \frac{r_3 - q}{r_3 - 1} \times
\begin{cases}
\langle \mathrm{dn}^2(\nu, k^2) \rangle & , 1 < q < r_2 \\
\langle \mathrm{cn}^2(\nu, k^{-2}) \rangle & , r_2 < q < r_3
\end{cases} .
\ee
The averages on the right-hand side of equation (\ref{eq:avg-cos2-gamma})
can be expressed as
\be
\label{eq:dn2-avg}
\langle \mathrm{dn}^2(\nu, k^2) \rangle = \frac{E(k^2)}{K(k^2)}
\ee
and
\be
\label{eq:cn2-avg}
\langle \mathrm{cn}^2(\nu, k^2) \rangle = \frac{\langle \mathrm{dn}^2(\nu, 
k^2) \rangle -1 + k^2}{k^2} .
\ee
We will employ this result in Section \ref{sec:external-processes}, where
we will also need the following results, all derived using equations
(\ref{eq:omega1-low-q})--(\ref{eq:cn2-avg}):
\be
\label{eq:sin-2-gamma-sin-2-alpha}
\langle \sin^2 \gamma \sin^2 \alpha \rangle =
\begin{cases}
\left( 1 - \langle \mathrm{dn}^2(\nu, k^2) \rangle \right)
\frac{r_3-q}{r_3 - r_2} & , 1 < q < r_2 \\
\left( 1 - \langle \mathrm{dn}^2(\nu, k^{-2}) \rangle \right)
\frac{q-1}{r_2 - 1} & , r_2 < q < r_3
\end{cases} , 
\ee
\be
\label{eq:sin-2-gamma-cos-2-alpha}
\langle \sin^2 \gamma \cos^2 \alpha \rangle = \frac{q - 1}{r_3 - 1} \times
\begin{cases}
\langle \mathrm{cn}^2(\nu, k^2) \rangle & , 1 < q < r_2 \\
\langle \mathrm{dn}^2(\nu, k^{-2}) \rangle & , r_2 < q < r_3
\end{cases} ,
\ee
\be
\label{eq:cos-gamma-av}
\langle \cos \gamma \rangle = \begin{cases}
\pm \left( \frac{r_3-q}{r_3-1}\right)^{1/2}
\frac{\upi}{2 K(k^2)} & , 1 < q < r_2 \\
0 & , r_2 < q < r_3
\end{cases} , 
\ee
\be
\label{eq:cos-alpha-sin-gamma-av}
\langle \sin \gamma \cos \alpha \rangle = \begin{cases}
0 & , 1 < q < r_2 \\
\pm \left( \frac{q-1}{r_3-1} \right)^{1/2} \frac{\upi}{2 K(k^{-2})} & , r_2 < q 
< r_3 
\end{cases} ,  
\ee
$\langle \sin \gamma \sin \alpha \rangle = 0$, 
$\langle \sin \gamma \cos \gamma \sin \alpha \rangle = 0$, 
$\langle \sin \gamma \cos \gamma \cos \alpha \rangle = 0$, and
$\langle \sin^2 \gamma \sin \alpha \cos \alpha \rangle = 0$. 
In equations (\ref{eq:cos-gamma-av}) and (\ref{eq:cos-alpha-sin-gamma-av}),
the + (-) signs are for the postive (negative) flip states with respect to
$\bmath{\hat{a}}_1$ and $\bmath{\hat{a}}_3$, respectively. 
From the definition of $q$ in equation (\ref{eq:q}), the expression 
$E = \frac{1}{2} \sum_i I_i \omega_i^2$, and equations
(\ref{eq:omega_1})--(\ref{eq:omega_3}), 
\be
q = \cos^2 \gamma + r_2 \sin^2 \gamma + r_3 \sin^2 \gamma \cos^2 \alpha .
\ee
Thus, for all values of $q$, 
\be
\label{eq:avg-quantity}
r_2 \langle \sin^2 \gamma \sin^2 \alpha \rangle + r_3 \langle \sin^2 \gamma 
\cos^2 \alpha \rangle = q - \langle \cos^2 \gamma \rangle .
\ee
This result will be useful in Section \ref{sec:external-processes}.


\section{External processes}

\label{sec:external-processes}

We assume that every gas-phase particle that strikes the grain returns to
the gas, via either thermal evaporation or incorporation into an H$_2$ molecule
that forms on the grain surface and is ejected. We take the arrival and
departure rates to be equal and apply a stochastic treatment for these
processes. Of the two torques associated with the interstellar magnetic field,
the Davis-Greenstein torque is treated deterministically and the Barnett
torque is omitted, since it only yields a precession in the angle
$\phi_B$, which is not relevant for any of the other dynamics under
consideration.

\subsection{Langevin equation}

Since the grain rotation time-scale is orders of magnitude smaller than all
other relevant time-scales, we average over grain rotation, assuming free
rotation, as described in Section \ref{sec:grain-rotation}, as a highly
accurate approximation. Thus, at any time $t$, the grain is characterized by
its angular momentum $\mathbfit{J}$, rotational kinetic energy $E$ or
its dimensionless measure $q$, and
flip state (with respect to either $\bmath{\hat{a}}_1$ or
$\bmath{\hat{a}}_3$, depending on the value of $q$). The time-scale for
internal relaxation (i.e.~Barnett plus nuclear) is
$\tau_{\mathrm{int}} = (\tau^{-1}_{\mathrm{Bar}} + \tau^{-1}_{\mathrm{nuc}})^{-1}$,
where the Barnett relaxation time-scale $\tau_{\mathrm{Bar}}$ is found from
equation (\ref{eq:tau-int-low-freq}) and the nuclear relaxation time-scale
is found using equations
(\ref{eq:tau-int-all-freq})--(\ref{eq:tau-int-low-freq}). Since
$\tau_{\mathrm{int}}$ is orders of magnitude shorter than the time-scales
associated with external processes, we neglect the role of external processes
in the evolution of the grain's rotational energy. That is, we simply evolve
$q$ using the Langevin equation (\ref{eq:internal-Langevin-eq}). 

The angular momentum $\mathbfit{J}$ must, of course, be tracked in
alignment coordinates, which are fixed in space (see Section
\ref{sec:coord-systems}). However, for the processes with stochastic
treatments, it is easier to evaluate the change
$\mathrm{d}\mathbfit{J}$ in instantaneous angular-momentum coordinates and then
transform the result to alignment coordinates. In this manner,
$\mathrm{d}\mathbfit{J}$ is found from three coupled Langevin equations:
\be
\mathrm{d}J_{i, J} = \langle \Gamma_{i, J}(\mathbfit{J}, q, \mathrm{fs}) \rangle
\, \mathrm{d}t + \sum_{j=1}^3 \langle B_{ij, J}(\mathbfit{J}, q, \mathrm{fs})
\rangle \, \mathrm{d}w_{j, J} \ \ \ \ (i=1-3) 
\ee
where $\mathrm{d}t$ is the time step and 
$\mathrm{d}w_{j, J}$ are Gaussian random variables with variance
$\mathrm{d}t$. 
The subscript `$J$' indicates that quantities are evaluated in 
angular-momentum coordinates, 
angle brackets denote averages over grain rotation, `fs' denotes the
flip state, which is positive or negative (equations \ref{eq:omega1-low-q},
\ref{eq:omega3-low-q}, \ref{eq:omega1-high-q}, \ref{eq:omega3-high-q}),
and $\langle B_{ij, J}(\mathbfit{J}, q, \mathrm{fs}) \rangle$ are components of
the matrix square root of the rotationally averaged diffusion tensor.
The components
$\langle \Gamma_{i, J}(\mathbfit{J}, q, \mathrm{fs}) \rangle$ of the rotationally
averaged mean torque and the components
$\langle C_{ij, J}(\mathbfit{J}, q, \mathrm{fs}) \rangle$ of the rotationally
averaged diffusion tensor can depend on $\mathbfit{J}$, $q$, and the flip
state. These quantities are evaluated in the following subsections.

\subsection{Collisions}

Suppose the gas, with temperature $T_{\mathrm{gas}}$, 
consists of particles with mass $m$ and number density $n$.
The gas thermal speed is defined as 
\be
\label{eq:v-th}
v_{\mathrm{th}} = \left( \frac{2 k_B T_{\mathrm{gas}}}{m} \right)^{1/2} 
\ee
where $k_B$ is Boltzmann's constant.
The velocity of a gas particle $\mathbfit{v} = v_{\mathrm{th}} \, \mathbfit{s}$. 
The `reduced velocity' $\mathbfit{s}$ is characterized
by polar angle $\theta_{\mathrm{in}}$ and azimuthal angle $\phi_{\mathrm{in}}$,
with $\bmath{\hat{N}}$ as the polar axis and $\bmath{\hat{t}}$ as the
azimuthal axis (recall equations \ref{eq:N-hat} and \ref{eq:t-hat}).  Thus,
\be
\label{eq:s-hat}
\bmath{\hat{s}} = - \left( \bmath{\hat{N}} \cos \theta_{\mathrm{in}} +
\bmath{\hat{t}} \sin 
\theta_{\mathrm{in}} \cos \phi_{\mathrm{in}} + \bmath{\hat{\phi}^{\prime}} \sin 
\theta_{\mathrm{in}} \sin \phi_{\mathrm{in}} \right) .
\ee
The Maxwell velocity distribution is 
\be
P(\mathbfit{s}) s^2 \, \mathrm{d}s \, \mathrm{d}\Omega = \upi^{-3/2} \exp(-s^2) 
s^2 \, \mathrm{d}s \, \mathrm{d}\Omega 
\ee
where $\mathrm{d}\Omega$ is the solid-angle element.

The velocity of the gas particle relative to a patch on the grain surface is
\be
\mathbfit{V} = v_{\mathrm{th}} s \, \bmath{\hat{s}} - \bmath{\omega}
\bmath{\times} \mathbfit{r}
\ee
where $\bmath{\omega}$ is the grain's angular velocity and $\mathbfit{r}$ is
the displacement from the grain's center of mass to the surface patch, with
components given by equations 
(\ref{eq:x-oblate-spheroidal})--(\ref{eq:z-oblate-spheroidal}). Thus, the 
rate at which gas particles with reduced speeds between $s$ and 
$s + \mathrm{d}s$ collide with a surface patch with area $\mathrm{d}S$ (equation
\ref{eq:surf-area-element}) from within solid angle 
$\mathrm{d}\Omega = \mathrm{d}(\cos \theta_{\mathrm{in}}) \, \mathrm{d}\phi_{\mathrm{in}}$ about the direction characterized by 
$(\theta_{\mathrm{in}}, \phi_{\mathrm{in}})$ is
\be
\label{eq:dR-col}
\mathrm{d}R_{\mathrm{col}} = \upi^{-3/2} n v_{\mathrm{th}} \, \mathrm{d}s \, s^2 
\exp(-s^2) \, \mathrm{d}(\cos \theta_{\mathrm{in}}) \, \mathrm{d}\phi_{\mathrm{in}} 
\, V^{\prime} \mathrm{d}S
\ee 
when $V^{\prime} > 0$ (and zero otherwise) where
\be
V^{\prime} = \left( s \bmath{\hat{s}} - \frac{\bmath{\omega} \bmath{\times}
\mathbfit{r}}{v_{\mathrm{th}}} \right) \bmath{\cdot} \left( - \bmath{\hat{N}}
\right) .
\ee
From equations (\ref{eq:x-oblate-spheroidal})--(\ref{eq:z-oblate-spheroidal}),
(\ref{eq:N-hat}), (\ref{eq:omega_1})--(\ref{eq:omega_3}), and 
(\ref{eq:s-hat}),
\begin{multline}
V^{\prime} = s \cos \theta_{\mathrm{in}} + \frac{J a_{\mathrm{eff}} \delta^{-1/3}}
{I_1 v_{\mathrm{th}}} \left( 1 - \delta^2 \right) A(\delta, \eta) \sin \eta
\cos \eta \sin \gamma \\
\times \left( r_2 \sin \phi^{\prime} \sin \alpha - r_3
\cos \phi^{\prime} \cos \alpha \right) ;
\label{eq:V-prime}
\end{multline}
$A(\delta, \eta)$ is defined in equation (\ref{eq:A-delta-eta}). 

We assume that $J a_{\mathrm{eff}}/I_1 v_{\mathrm{th}} \ll 1$, so the second
term in the expression for $V^{\prime}$ can be neglected unless the first
term yields a zero integral and the lower limit in integrals over $s$ can
simply be taken to be zero. 

The angular momentum acquired by the grain when a gas-phase particle collides
and sticks to the surface, as observed in an inertial frame, is
\be
\label{eq:Delta-J-col}
\bmath{\Delta J}_{\mathrm{col}} = m v_{\mathrm{th}} s \, \mathbfit{r} \bmath{\times} 
\bmath{\hat{s}} = m v_{\mathrm{th}} a_{\mathrm{eff}} \delta^{-1/3} 
\bmath{\Delta J^{\prime}}_{\mathrm{col}} ,
\ee
where, from equations 
(\ref{eq:x-oblate-spheroidal})--(\ref{eq:z-oblate-spheroidal}), 
(\ref{eq:N-hat})--(\ref{eq:t-hat}), and (\ref{eq:s-hat}),
\begin{multline}
\Delta J^{\prime}_{x, \mathrm{col}} = s \left[ - A(\delta, \eta) \left( 1 - 
\delta^2 \right) \sin \eta \cos \eta \sin \phi^{\prime} \cos \theta_{\mathrm{in}}
\right. \\
+ A(\delta, \eta) \delta \sin \phi^{\prime} \sin \theta_{\mathrm{in}} \cos
\phi_{\mathrm{in}} + \delta \sin \eta \cos \phi^{\prime} \sin \theta_{\mathrm{in}}
\sin \phi_{\mathrm{in}}  \Big] ,
\label{eq:Delta-J-prime-x}
\end{multline}
\begin{multline}
\Delta J^{\prime}_{y, \mathrm{col}} = s \left[ A(\delta, \eta) \left( 1 - 
\delta^2 \right) \sin \eta \cos \eta \cos \phi^{\prime} \cos \theta_{\mathrm{in}}
\right. \\
- A(\delta, \eta) \delta \cos \phi^{\prime} \sin \theta_{\mathrm{in}} \cos
\phi_{\mathrm{in}} + \delta \sin \eta \sin \phi^{\prime} \sin \theta_{\mathrm{in}}
\sin \phi_{\mathrm{in}}  \Big] ,
\label{eq:Delta-J-prime-y}
\end{multline}
\be
\label{eq:Delta-J-prime-z}
\Delta J^{\prime}_{z, \mathrm{col}} = - s \cos \eta \sin \theta_{\mathrm{in}}
\sin \phi_{\mathrm{in}} .
\ee

The mean torque on the grain due to collisions with gas particles is
\be
\label{eq:Gamma-col-general}
\bmath{\Gamma}_{\mathrm{col}} = \int \mathrm{d}R_{\mathrm{col}} \, \bmath{\Delta
J}_{\mathrm{col}} .
\ee
From equations (\ref{eq:dR-col}), (\ref{eq:V-prime}), 
(\ref{eq:Delta-J-col})--(\ref{eq:Delta-J-prime-z}), and
(\ref{eq:Gamma-col-general}), 
\begin{multline}
\bmath{\Gamma}_{\mathrm{col}} = - \frac{\sqrt{\upi}}{2} n m v_{\mathrm{th}}
a_{\mathrm{eff}}^4 \delta^{-4/3} \left( 1 - \delta^2 \right)^2 
\mathcal{I}_4(\delta) \frac{J}{I_1} \sin \gamma \\
\times \left( r_2 \sin \alpha \,
\bmath{\hat{x}} + r_3 \cos \alpha \bmath{\hat{y}} \right) .
\end{multline}
From equations (\ref{eq:x-hat}), (\ref{eq:y-hat}), and 
(\ref{eq:avg-grain-rot}), the rotationally averaged torque is
\begin{multline}
\langle \bmath{\Gamma}_{\mathrm{col}} \rangle = - \frac{\sqrt{\upi}}{2} mn
v_{\mathrm{th}} a_{\mathrm{eff}}^4 \delta^{-4/3} \left( 1 - \delta^2 \right)^2  
\mathcal{I}_4(\delta) \, \frac{\mathbfit{J}}{I_1} \\
\times \left( r_2 \langle 
\sin^2 \gamma \sin^2 \alpha \rangle + r_3 \langle \sin^2 \gamma \cos^2 \alpha 
\rangle \right) ;
\label{eq:Gamma-col-avg}
\end{multline}
expressions for $\mathcal{I}_4(\delta)$ and
$r_2 \langle \sin^2 \gamma \sin^2 \alpha \rangle + r_3 \langle \sin^2 \gamma 
\cos^2 \alpha \rangle$ are given in equations (\ref{eq:app-i4}) and 
(\ref{eq:avg-quantity}). 
Note that $\langle \bmath{\Gamma}_{\mathrm{col}} \rangle = 0$ for steady 
rotation about the geometric symmetry axis (i.e.~$\gamma = 0$), in 
agreement with previous results \citep{PS71, RDF93}. 

The diffusion coefficients are given by
\be
C_{ij, \mathrm{col}} = \int \mathrm{d}R_{\mathrm{col}} \, \Delta J_{i, \mathrm{col}}
\Delta J_{j, \mathrm{col}} .
\ee
The diffusion tensor in grain-body coordinates is diagonal, with
\be
\label{eq:C_zz-body}
C_{zz, \mathrm{col}} = \frac{2 \sqrt{\upi}}{3} nm^2
v_{\mathrm{th}}^3 a_{\mathrm{eff}}^4 \delta^{-4/3} Z_1(\delta) ,
\ee
\be
\label{eq:C_xx-body}
C_{xx, \mathrm{col}} = C_{yy, \mathrm{col}} = \frac{2 \sqrt{\upi}}{3} nm^2
v_{\mathrm{th}}^3 a_{\mathrm{eff}}^4 \delta^{-4/3} Z_2(\delta) ,
\ee
where
\be
\label{eq:Z-1}
Z_1(\delta) = \frac{3}{4} \left[ \mathcal{I}_2(\delta) - \mathcal{I}_1(\delta)
\right] = \frac{3}{16} \left[ 3 + 4 \delta^2 g(\delta) - \frac{1 - \delta^4
g(\delta)}{1 - \delta^2} \right]
\ee
and
\begin{multline}
Z_2(\delta) = \frac{3}{8} \left\{ 2 \left( 1 - \delta^2 \right)^2
\mathcal{I}_4(\delta) + \delta^2 \left[ \mathcal{I}_1(\delta) +
\mathcal{I}_3(\delta) \right] \right\} \\
= \frac{3}{32} \ \frac{4 - 3 \delta^4 + \delta^4 (2 - 3 \delta^2) g(\delta)}
{1 - \delta^2} .
\label{eq:Z-2}
\end{multline}
The integrals $\mathcal{I}_i(\delta)$ and the function $g(\delta)$ are 
defined in Section \ref{sec:spheroid-properties}.
Note that, to within the approximations adopted here, $C_{zz, \mathrm{col}}$ and
$C_{xx, \mathrm{col}}$ are independent of the grain rotation and that the results
in equations (\ref{eq:C_zz-body}) and (\ref{eq:C_xx-body}) are identical to 
those found by \citet{RDF93}. 

Transforming the diffusion tensor to angular-momentum coordinates and
averaging over grain rotation,
\be
\label{eq:C-zz-J-avg}
\langle C_{zz, J, \mathrm{col}} \rangle = \frac{2 \sqrt{\upi}}{3} nm^2 
v_{\mathrm{th}}^3 
a_{\mathrm{eff}}^4 \delta^{-4/3} \left[ Z_1(\delta) \langle \cos^2 \gamma
\rangle + Z_2(\delta) \langle \sin^2 \gamma \rangle \right] 
\ee
and
\begin{multline}
\langle C_{xx, J, \mathrm{col}} \rangle = \langle C_{yy, J, \mathrm{col}} \rangle =
\frac{2 
\sqrt{\upi}}{3} nm^2 v_{\mathrm{th}}^3 a_{\mathrm{eff}}^4 \delta^{-4/3} \\
\times \frac{1}{2} 
\left[ Z_2(\delta) \left( 1 + \langle \cos^2 \gamma \rangle \right) 
+ Z_1(\delta) \langle \sin^2 \gamma \rangle \right] .
\label{eq:C-xx-J-avg}
\end{multline}

\subsection{Evaporation}

\label{sec:evap}

Consider thermal evaporation of particles of mass $m_{\mathrm{ev}}$, distributed
uniformly across the grain 
surface. The total evaporation rate must equal the total collision
rate and detailed balancing applies when the evaporation temperature
$T_{\mathrm{ev}}$ equals the gas temperature $T_{\mathrm{gas}}$. Thus, as described
in Appendix B in \citet{RDF93}, the rate at which particles evaporate from a 
surface patch with area $\mathrm{d}S$, with speeds between
$v_{\mathrm{th, ev}} \, s$ and $v_{\mathrm{th, ev}} (s + \mathrm{d}s)$ and from within
solid angle 
$\mathrm{d}\Omega = \mathrm{d}(\cos \theta_{\mathrm{in}}) \mathrm{d}\phi_{\mathrm{in}}$ 
about the direction characterized by $(\theta_{\mathrm{in}}, \phi_{\mathrm{in}})$, is
\be
\label{eq:dR-evap}
\mathrm{d}R_{\mathrm{ev}} = \upi^{-3/2}  \frac{m}{m_{\mathrm{ev}}} n v_{\mathrm{th}} \, 
\mathrm{d}s \, s^3 \exp(-s^2) \, \mathrm{d}(\cos \theta_{\mathrm{in}}) \, \cos 
\theta_{\mathrm{in}} \, \mathrm{d}\phi_{\mathrm{in}} \, \mathrm{d}S .
\ee
The evaporative thermal speed $v_{\mathrm{th, ev}}$ is defined identically to
the gas thermal speed, except $T_{\mathrm{gas}}$ and $m$ in equation
(\ref{eq:v-th}) are replaced with $T_{\mathrm{ev}}$ and $m_{\mathrm{ev}}$. 

The angular momentum imparted to the grain following an evaporation event
is
\be
\label{eq:Delta-J-evap}
\bmath{\Delta J}_{\mathrm{ev}} = m_{\mathrm{ev}} \mathbfit{r} \bmath{\times} \left( 
v_{\mathrm{th, ev}} \, s \, \bmath{\hat{s}} - \bmath{\omega} \bmath{\times}
\mathbfit{r} \right) = m v_{\mathrm{th}} a_{\mathrm{eff}} \delta^{-1/3} 
\bmath{\Delta J^{\prime}}_{\mathrm{ev}}
\ee
where, from equations 
(\ref{eq:x-oblate-spheroidal})--(\ref{eq:z-oblate-spheroidal}),
(\ref{eq:N-hat})--(\ref{eq:t-hat}), (\ref{eq:omega_1})--(\ref{eq:omega_3}), and 
(\ref{eq:s-hat}),
\be
\bmath{\Delta J^{\prime}}_{\mathrm{ev}} = \bmath{\Delta J^{\prime}}_{\mathrm{ev}}(1) + 
\bmath{\Delta J^{\prime}}_{\mathrm{ev}}(2) ,
\ee
\be
\label{eq:Delta-J-prime-ev-1}
\bmath{\Delta J^{\prime}}_{\mathrm{ev}}(1) = \frac{m_{\mathrm{ev}} v_{\mathrm{th, ev}}}
{m v_{\mathrm{th}}} \, \bmath{\Delta J^{\prime}}_{\mathrm{col}} ,
\ee
\begin{multline}
\Delta J^{\prime}_{x, \mathrm{ev}}(2) = \frac{m_{\mathrm{ev}}}{m} \ \frac{J
a_{\mathrm{eff}} \delta^{-1/3}}{I_1 v_{\mathrm{th}}} \left( r_3 \sin \gamma \cos
\alpha \cos^2 \eta \sin \phi^{\prime} \cos \phi^{\prime} \right. \\
- r_2 \sin \gamma \sin
\alpha \cos^2 \eta \sin^2 \phi^{\prime} - \delta^2 r_2 \sin \gamma \sin \alpha
\sin^2 \eta \\
+ \delta \cos \gamma \sin \eta \cos \eta \cos \phi^{\prime} \Big) \, ,
\label{eq:Delta-J-prime-x-ev-2}
\end{multline}
\begin{multline}
\Delta J^{\prime}_{y, \mathrm{ev}}(2) = \frac{m_{\mathrm{ev}}}{m} \ \frac{J
a_{\mathrm{eff}} \delta^{-1/3}}{I_1 v_{\mathrm{th}}} \left( r_2 \sin \gamma \sin
\alpha \cos^2 \eta \sin \phi^{\prime} \cos \phi^{\prime} \right. \\
- r_3 \sin \gamma \cos
\alpha \cos^2 \eta \cos^2 \phi^{\prime} - \delta^2 r_3 \sin \gamma \cos \alpha
\sin^2 \eta \\
+ \delta \cos \gamma \sin \eta \cos \eta \sin \phi^{\prime} \Big) \, ,
\end{multline}
\begin{multline}
\Delta J^{\prime}_{z, \mathrm{ev}}(2) = \frac{m_{\mathrm{ev}}}{m} \ \frac{J
a_{\mathrm{eff}} \delta^{-1/3}}{I_1 v_{\mathrm{th}}} \Big[ \delta \sin \gamma \sin
\eta \cos \eta \\
\left. \times \left( r_2 \sin \alpha \cos \phi^{\prime} + r_3 \cos \alpha
\sin \phi^{\prime} \right) - \cos \gamma \cos^2 \eta \right] .
\label{eq:Delta-J-prime-z-ev-2}
\end{multline}

Evaluation of the mean torque and diffusion tensor associated with evaporation
proceeds in the same way as for collisions, yielding
\begin{multline}
\langle \bmath{\Gamma}_{\mathrm{ev}} \rangle = - \frac{\sqrt{\upi}}{2} mn
v_{\mathrm{th}} a_{\mathrm{eff}}^4 \delta^{-4/3} \frac{\mathbfit{J}}{I_1}
\left\{ 2 \left[ \mathcal{I}_2(
\delta) - \mathcal{I}_1(\delta) \right] \langle \cos^2 \gamma \rangle +
\right. \\
\left. \left[ \mathcal{I}_2(\delta) - \left( 1 - 2 \delta^2
\right) \mathcal{I}_1(\delta) \right] \left( r_2 \langle \sin^2 \gamma 
\sin^2 \alpha \rangle + r_3 \langle \sin^2 \gamma \cos^2 \alpha
\rangle \right) \right\} 
\label{eq:Gamma-ev-avg}
\end{multline}
and
\be
\label{eq:C-ij-ev}
C_{ij, \mathrm{ev}} = \frac{T_{\mathrm{ev}}}{T_{\mathrm{gas}}} \, C_{ij, \mathrm{col}} .
\ee
The prefactor in equation (\ref{eq:C-ij-ev}) results as follows: 
$C_{ij, \mathrm{col}} \propto m^2 v_{\mathrm{th}}^3$ and 
$C_{ij, \mathrm{ev}} \propto m m_{\mathrm{ev}} v_{\mathrm{th}} v_{\mathrm{th, ev}}^2$,
so $C_{ij, \mathrm{ev}}/C_{ij, \mathrm{col}} = m_{\mathrm{ev}} v_{\mathrm{th, ev}}^2 /
(m v_{\mathrm{th}}^2) = T_{\mathrm{ev}}/T_{\mathrm{gas}}$. 
When $r_2 = r_3$ and $\gamma = 0$, equations (\ref{eq:Gamma-ev-avg}) and
(\ref{eq:C-ij-ev}) reduce to equations C17 and C21 in \citet{RDF93} for 
the mean torque and diffusion tensor for an oblate spheroid with
dynamic symmetry rotating steadily about the symmetry axis. 

\subsection{Formation and ejection of H$_2$}

Now consider the case that particles depart the grain as newly formed 
H$_2$ molecules. For simplicity, we will assume that the molecules have a 
fixed kinetic energy $E_{\mathrm{H}2}$. With $m_{\mathrm{H}2} = 2 m$, the
departure speed is $v_{\mathrm{H}2} = (2 E_{\mathrm{H}2}/m_{\mathrm{H}2})^{1/2}$. 
For convenience, define $T_{\mathrm{H}2} = E_{\mathrm{H2}}/k_B$. 

\subsubsection{Uniformly distributed formation sites}
\label{sec:uniform-sites}

First suppose that the formation sites fully cover the grain surface, with
a fixed number density per unit surface area. If the departing molecules
are distributed uniformly in solid angle, then the analysis in Section
\ref{sec:evap} applies with minor modification. Integrating equation
(\ref{eq:dR-evap}) over $s$, the departure rate from a surface patch is
\be
\label{eq:dR-H2}
\mathrm{d}R_{\mathrm{H}2} = \frac{\upi^{-3/2}}{2}  \frac{m}{m_{\mathrm{H}2}} n 
v_{\mathrm{th}} \, \mathrm{d}(\cos \theta_{\mathrm{in}}) \, \cos \theta_{\mathrm{in}} \,
\mathrm{d}\phi_{\mathrm{in}} \, \mathrm{d}S .
\ee
The angular momentum imparted to the grain when an H$_2$ molecule departs
is identical to that in equation (\ref{eq:Delta-J-evap}) except that
$m_{\mathrm{ev}}$ is replaced with $m_{\mathrm{H}2}$ and 
$v_{\mathrm{th, ev}} \, s$ is replaced with $v_{\mathrm{H}2}$. Thus,
\be
\bmath{\Delta J^{\prime}}_{\mathrm{H}2} = \bmath{\Delta J^{\prime}}_{\mathrm{H2}}(1) + 
\bmath{\Delta J^{\prime}}_{\mathrm{H2}}(2) ,
\ee
where the components of $\bmath{\Delta J^{\prime}}_{\mathrm{H2}}(2)$ are given by
equations (\ref{eq:Delta-J-prime-x-ev-2})--(\ref{eq:Delta-J-prime-z-ev-2}),
except with $m_{\mathrm{ev}}$ replaced by $m_{\mathrm{H}2}$, and
$\bmath{\Delta J^{\prime}}_{\mathrm{H2}}(1)$ is given by equation 
(\ref{eq:Delta-J-prime-ev-1}), except with 
$m_{\mathrm{ev}} v_{\mathrm{th, ev}}$ replaced by $m_{\mathrm{H}2} \, v_{\mathrm{H}2}/s$.
In evaluating the mean torque, the term involving 
$\bmath{\Delta J^{\prime}}_{\mathrm{H2}}(1)$ vanishes upon integration. Thus,
\be
\label{eq:Gamma-H2}
\langle \bmath{\Gamma}_{\mathrm{H}2} \rangle = \langle \bmath{\Gamma}_{\mathrm{ev}} 
\rangle .
\ee
With the assumption that $J a_{\mathrm{eff}}/I_1 v_{\mathrm{th}} \ll 1$, only 
the term involving $\bmath{\Delta J^{\prime}}_{\mathrm{H2}}(1)$ contributes to
the diffusion tensor. In the calculation of $C_{ij, \mathrm{ev}}$, a term
$\int_0^{\infty} ds \, s^5 \exp(-s^2)$ arises. This term evaluates to unity
and is replaced by the factor $\frac{1}{2}$ in equation (\ref{eq:dR-H2})
in the calculation of $C_{ij, \mathrm{H}2}$.
Thus, 
\be
\frac{C_{ij, \mathrm{H}2}}{C_{ij, \mathrm{ev}}} = \frac{\frac{1}{2} m
m_{\mathrm{H}2} v_{\mathrm{th}} v_{\mathrm{H}2}^2}{m m_{\mathrm{ev}} v_{\mathrm{th}}
v_{\mathrm{th, ev}}^2} .
\ee
Finally, 
\be
\label{eq:C-ij-H2}
C_{ij, \mathrm{H}2} = \frac{1}{2} \, \frac{T_{\mathrm{H}2}}{T_{\mathrm{ev}}} \  
C_{ij, \mathrm{ev}} = \frac{1}{2} \,  
\frac{T_{\mathrm{H}2}}{T_{\mathrm{gas}}} \  C_{ij, \mathrm{col}} .
\ee

\subsubsection{Special formation sites}

\label{sec:special-sites}

Now consider a grain where H$_2$ formation only occurs at a set of $N_s$
special surface sites. We randomly select the position 
$(\eta_i, \phi^{\prime}_i)$ of
each site from a uniform distribution (in surface area) over the surface. 
From equation (\ref{eq:surf-area-element}), the surface area element is
\be
\mathrm{d}S = a_{\mathrm{eff}}^2 \delta^{-2/3} \mathrm{d}u \, \mathrm{d}\phi^{\prime}
\ee
with
\be
\mathrm{d}u = \left[ \delta^2 + (1-\delta^2) 
\sin^2 \eta \right]^{1/2} \cos \eta \, \mathrm{d}\eta .
\ee
Thus, for each site $\phi^{\prime}_i$ is selected randomly from a uniform 
distribution in $\phi^{\prime}$ (0 to $2 \upi$) and $\eta_i$ is selected
randomly from a uniform distribution in $u$, where $u$ and $\eta$ are 
related by
\begin{multline}
u = \frac{1}{2} \left[ 1 + \sin \eta \sqrt{\delta^2 + \left(1-\delta^2 
\right) \sin^2 \eta} \right] 
+ \frac{\delta^2}{2 \sqrt{1-\delta^2}} \\
\times \left\{ \ln \left[
\left( 1 + \sqrt{1-\delta^2} \right) \left( \sqrt{1-\delta^2} \sin \eta + 
\sqrt{\delta^2 + \left(1-\delta^2 \right) \sin^2 \eta} \right) \right] \right.
\\
- 2 \ln \delta \bigg\} .
\end{multline}
As $\eta$ ranges from $-\upi/2$ to $\upi/2$, $u$ ranges from 0 to 
\be
u_{\mathrm{max}} = 1 + \frac{\delta^2}{\sqrt{1-\delta^2}} \ln \left( \frac{1 + 
\sqrt{1-\delta^2}}{\delta} \right) = 1 + \delta^2 g(\delta) .
\ee

We will assume that molecule formation occurs at the same rate at each
surface site. Thus, the rate per site at which molecules are ejected is
\be
R_{\mathrm{H}2}(\mathrm{per \ site}) = \frac{R_{\mathrm{col}}}{2 N_s} ,
\ee
where $R_{\mathrm{col}}$ is the total rate at which gas atoms collide with the
grain and the factor of $\frac{1}{2}$ appears since there are 2 H atoms per
H$_2$ molecule. From equations (\ref{eq:dR-col}), (\ref{eq:V-prime}),
(\ref{eq:surf-area-element}), and (\ref{eq:app-i2}), 
\be
R_{\mathrm{col}} = \sqrt{\upi} n v_{\mathrm{th}} a_{\mathrm{eff}}^2 \delta^{-2/3} \left[ 1 
+ \delta^2 g(\delta) \right] .
\ee

A systematic torque only has the potential to maintain suprathermal grain
rotation if it has a non-zero component along 
$\bmath{\hat{z}} = \bmath{\hat{a}}_1$. Otherwise, from equations
(\ref{eq:x-hat}) and (\ref{eq:y-hat}), the component of the rotationally 
averaged systematic
torque along $\bmath{\hat{J}} = \bmath{\hat{z}}_J$ vanishes in the limit
$\gamma \rightarrow 0$, which characterizes suprathermal rotation. 
From equation (\ref{eq:Delta-J-prime-z}), the component of the torque along
$\bmath{\hat{z}}$ vanishes if the outgoing molecules are uniformly 
distributed in solid angle, or if they depart along the surface normal
$\bmath{\hat{N}}$. This is a consequence of the spheroidal shape, for which
the components of $\mathbfit{r}$ (the displacement from the grain center to
the surface patch) and $\bmath{\hat{N}}$ that are
perpendicular to $\bmath{\hat{z}}$ lie in the same direction, so that 
$\mathbfit{r} \bmath{\times} \bmath{\hat{N}}$ has zero component along 
$\bmath{\hat{z}}$.  
So rather than taking the outgoing molecules to be uniformly distributed in
solid angle, we randomly pick angles $\theta_{\mathrm{out}, i}$ (from a 
uniform distribution in $\cos^2 \theta_{\mathrm{out}}$ between  
$[(\cos \theta_{\mathrm{out}})_{\mathrm{min}}]^2$ and 1) and
$\phi_{\mathrm{out}, i}$ (uniformly distributed between 0 and $2 \upi$) for each
site such that a molecule departing site $i$ has velocity 
\begin{multline}
\mathbfit{v}_{\mathrm{H}2} = v_{\mathrm{H}2} \left( \bmath{\hat{N}}_i \cos 
\theta_{\mathrm{out}, i} + \bmath{\hat{t}}_i \sin \theta_{\mathrm{out}, i} \cos 
\phi_{\mathrm{out}, i} \right. \\
\left. + \bmath{\hat{\phi}^{\prime}}_i \sin \theta_{\mathrm{out}, i} 
\sin \phi_{\mathrm{out}, i} \right) .
\label{eq:v-H2-special-sites}
\end{multline}
The form of the distribution for $\theta_{\mathrm{out}, i}$ was chosen for
consistency with the treatment of the case with uniformly distributed 
formation sites. In our simulations, we adopt 
$(\cos \theta_{\mathrm{out}})_{\mathrm{min}} = 0.8$. 

In contrast to the case where the surface is uniformly covered with 
H$_2$-formation sites, the mean torque associated with 
$\bmath{\Delta J^{\prime}}_{\mathrm{H}2}(1)$ does not vanish. This term accounts
for the systematic torque, while the term associated with 
$\bmath{\Delta J^{\prime}}_{\mathrm{H}2}(2)$ combines with the torque due to
collisions with atoms to account for the drag torque. The mean systematic
torque is
\be
\bmath{\Gamma}_{\mathrm{H2, \, sys}} = R_{\mathrm{H2}}(\mathrm{per \ site}) 
\sum_{i=1}^{N_s} \bmath{\Delta J}_{\mathrm{H2}}(1)_{i} .
\ee
The rotationally averaged systematic torque is 
\begin{multline}
\langle \mathbf{\Gamma}_{\mathrm{H2, \, sys}} \rangle = \sqrt{\upi} m n 
v_{\mathrm{th}} v_{\mathrm{H2}} a^3_{\mathrm{eff}} \delta^{-1} \left[ 1 + \delta^2
g(\delta) \right] \\
\times \left( Q_1 \langle \cos \gamma \rangle + Q_2
\langle \sin \gamma \cos \alpha \rangle \right) \bmath{\hat{J}} ,
\label{eq:Gamma-H2-sys-rot-avg}
\end{multline}
where 
\be
Q_1 = - \frac{1}{N_s} \sum_{i=1}^{N_s} \cos \eta_i \sin \theta_{\mathrm{out}, i}
\sin \phi_{\mathrm{out}, i} ,
\ee
\begin{multline}
Q_2 = \frac{1}{N_s} \sum_{i=1}^{N_s} \left[ \left( 1 - \delta^2 \right) 
A(\delta, \eta_i) \sin \eta_i \cos \eta_i \cos \phi^{\prime}_i \cos
\theta_{\mathrm{out}, i} \right. \\
- \delta A(\delta, \eta_i) \cos \phi^{\prime}_i 
\sin \theta_{\mathrm{out}, i} \cos \phi_{\mathrm{out}, i} \\
+ \delta \sin \eta_i
\sin \phi^{\prime}_i \sin \theta_{\mathrm{out},i} \sin \phi_{\mathrm{out},i} \Big] ,
\end{multline}
and $\langle \cos \gamma \rangle$ and 
$\langle \sin \gamma \cos \alpha \rangle$ are evaluated in equations
(\ref{eq:cos-gamma-av}) and  (\ref{eq:cos-alpha-sin-gamma-av}).

The contribution to the drag torque is evaluated similarly, except that
$\bmath{\Delta J}_{\mathrm{H2}}(1)$ is replaced with 
$\bmath{\Delta J}_{\mathrm{H2}}(2)$. After some algebra, we find that
\begin{multline}
\langle \mathbf{\Gamma}_{\mathrm{H2, \, drag}} \rangle = - \sqrt{\upi} m n 
v_{\mathrm{th}} a^4_{\mathrm{eff}} \delta^{-4/3} \left[ 1 + \delta^2
g(\delta) \right] \frac{\mathbfit{J}}{I_1} \\
\times \left( Q_3 \langle \cos^2 \gamma 
\rangle + Q_4 \langle \sin^2 \gamma \sin^2 \alpha \rangle + 
Q_5 \langle \sin^2 \gamma \cos^2 \alpha \rangle \right) ,
\label{eq:Gamma-H2-drag}
\end{multline}
where
\be
Q_3 = \frac{1}{N_s} \sum_{i=1}^{N_s} \cos^2 \eta_i , 
\ee
\be
Q_4 = \frac{r_2}{N_s} \sum_{i=1}^{N_s} \left( \cos^2 \eta_i \sin^2 \phi^{\prime}_i
+ \delta^2 \sin^2 \eta_i \right) ,
\ee
\be
Q_5 = \frac{r_3}{N_s} \sum_{i=1}^{N_s} \left( \cos^2 \eta_i \cos^2 \phi^{\prime}_i
+ \delta^2 \sin^2 \eta_i \right) ,
\ee
and $\langle \cos^2 \gamma \rangle$, 
$\langle \sin^2 \gamma \sin^2 \alpha \rangle$,
and $\langle \sin^2 \gamma \cos^2 \alpha \rangle$ are evaluated in 
equations (\ref{eq:avg-cos2-gamma}), (\ref{eq:sin-2-gamma-sin-2-alpha}),
and (\ref{eq:sin-2-gamma-cos-2-alpha}). 
In the limit of uniform surface coverage of H$_2$-formation sites, equation
(\ref{eq:Gamma-H2-drag}) reproduces equation (\ref{eq:Gamma-H2}). 

The diffusion tensor in grain-body coordinates is given by
\begin{multline}
C_{ij, \mathrm{H}2} = 2 \sqrt{\upi} n m^2 v_{\mathrm{th}} v^2_{\mathrm{H}2} 
a^4_{\mathrm{eff}} \delta^{-4/3} \left[ 1 + \delta^2 g \left( \delta \right)
\right] \\
\times \frac{1}{N_s} \sum_{\rho = 1}^{N_s} \Delta 
\tilde{J}_{\mathrm{H}2, i}(1)_{\rho} \Delta \tilde{J}_{\mathrm{H}2, j}(1)_{\rho} ,
\label{eq:C-ij-H2-special-sites}
\end{multline}
where
\begin{multline}
\Delta \tilde{J}_{\mathrm{H}2, x}(1)_{\rho} = - A(\delta, \eta_{\rho}) \left( 1 - 
\delta^2 \right) \sin \eta_{\rho} \cos \eta_{\rho} \sin \phi_{\rho}^{\prime} \cos 
\theta_{\mathrm{out}, \rho} \\
+ A(\delta, \eta_{\rho}) \delta \sin \phi_{\rho}^{\prime} 
\sin \theta_{\mathrm{out}, \rho} \cos \phi_{\mathrm{out}, \rho} \\
+ \delta \sin \eta_{\rho} 
\cos \phi_{\rho}^{\prime} \sin \theta_{\mathrm{out}, \rho} \sin \phi_{\mathrm{out}, \rho} ,
\label{eq:Delta-J-tilde-x}
\end{multline}
\begin{multline}
\Delta \tilde{J}_{\mathrm{H}2, y}(1)_{\rho} = A(\delta, \eta_{\rho}) \left( 1 - 
\delta^2 \right) \sin \eta_{\rho} \cos \eta_{\rho} \cos \phi_{\rho}^{\prime} \cos 
\theta_{\mathrm{out}, \rho} \\
- A(\delta, \eta_{\rho}) \delta \cos \phi_{\rho}^{\prime} 
\sin \theta_{\mathrm{out}, \rho} \cos \phi_{\mathrm{out}, \rho} \\
+ \delta \sin \eta_{\rho} 
\sin \phi_{\rho}^{\prime} \sin \theta_{\mathrm{out}, \rho} \sin \phi_{\mathrm{out}, \rho} ,
\label{eq:Delta-J-tilde-y}
\end{multline}
\be
\label{eq:Delta-J-tilde-z}
\Delta \tilde{J}_{\mathrm{H}2, z}(1)_{\rho} = - \cos \eta_{\rho} \sin 
\theta_{\mathrm{out}, \rho} \sin \phi_{\mathrm{out}, \rho} .
\ee
Transforming to angular-momentum coordinates and averaging over grain rotation,
\begin{multline}
\langle C_{xx, J, \mathrm{H}2} \rangle = \langle C_{yy, J, \mathrm{H}2} \rangle =
\frac{1}{2} C_{xx, \mathrm{H}2} \left( 1 - \langle \sin^2 \gamma \sin^2 \alpha
\rangle \right) \\
+ \frac{1}{2} C_{yy, \mathrm{H}2} \left( 1 - \langle \sin^2 \gamma 
\cos^2 \alpha \rangle \right) + \frac{1}{2} C_{zz, \mathrm{H}2} \langle \sin^2
\gamma \rangle ,
\label{eq:C-xx-J-H2-sites}
\end{multline}
\begin{multline}
\langle C_{zz, J, \mathrm{H}2} \rangle = C_{xx, \mathrm{H}2} \langle \sin^2 \gamma 
\sin^2 \alpha \rangle + C_{yy, \mathrm{H}2} \langle \sin^2 \gamma \cos^2 \alpha 
\rangle \\
+ C_{zz, \mathrm{H}2} \langle \cos^2 \gamma \rangle ,
\label{eq:C-zz-J-H2-sites}
\end{multline}
and the off-diagonal elements all vanish. If 
$C_{xx, \mathrm{H}2} = C_{yy, \mathrm{H}2}$, then equations (\ref{eq:C-xx-J-H2-sites})
and (\ref{eq:C-zz-J-H2-sites}) adopt the form of equations 
(\ref{eq:C-xx-J-avg}) and (\ref{eq:C-zz-J-avg}). 
Since only the three diagonal element of $C_{ij, \mathrm{H}2}$ are needed, we
define three additional dimensionless efficiency factors,
\be
Q_6 = \frac{1}{N_s} \sum_{\rho = 1}^{N_s} \left[ \Delta \tilde{J}_{\mathrm{H}2, x}
(1)_{\rho} \right]^2 ,
\label{eq:Q_6-def}
\ee
\be
Q_7 = \frac{1}{N_s} \sum_{\rho = 1}^{N_s} \left[ \Delta \tilde{J}_{\mathrm{H}2, y}
(1)_{\rho} \right]^2 ,
\label{eq:Q_7-def}
\ee
\be
Q_8 = \frac{1}{N_s} \sum_{\rho = 1}^{N_s} \left[ \Delta \tilde{J}_{\mathrm{H}2, z}
(1)_{\rho} \right]^2 .
\label{eq:Q_8-def}
\ee
From equation (\ref{eq:C-ij-H2-special-sites}),
$C_{xx, \mathrm{H}2} \propto Q_6$,
$C_{yy, \mathrm{H}2} \propto Q_7$, and
$C_{zz, \mathrm{H}2} \propto Q_8$.

\subsection{Davis-Greenstein Torque} \label{sec:D-G}

\citet{DG51} evaluated the rotation-averaged torque due to paramagnetic
dissipation in the interstellar magnetic field for the case of an oblate grain
with dynamic symmetry:
\be
\label{eq:DG-torque}
\langle \bmath{\Gamma}_{\mathrm{DG}} \rangle = -
\tau_{\mathrm{DG}}^{-1} \left[ 1 + \left( r_2 - 1 \right) \sin^2 \gamma \right]
\left( J_{x, B} \, \bmath{\hat{x}}_B + J_{y,B} \bmath{\hat{y}}_B \right) ,
\ee
where the Davis-Greenstein timescale is
\begin{multline}
\tau_{\mathrm{DG}} = \frac{2 \alpha_1 \bar{\rho} a_{\mathrm{eff}}^2}{5 \chi_0 T_2 B^2}
\\
\approx 1.52 \times 10^6 \, \mathrm{yr} \left( \frac{\alpha_1 \bar{\rho}}
{3 \, \mathrm{g} \, \mathrm{cm}^{-3}} \right) \left( \frac{a_{\mathrm{eff}}}
{0.1 \, \mu \mathrm{m}} \right)^2
\left( \frac{T_d}{15 \, \mathrm{K}} \right) \left( \frac{B}{5 \mu \mathrm{G}} 
\right)^{-2} .
\label{eq:tau-DG}
\end{multline}
We assume that for electron paramagnetism, 
$\chi^{\prime \prime}/\omega = \chi_0 T_2 = 10^{-13} (T_d/15 \, \mathrm{K})^{-1}
\, \mathrm{s}$, where $\chi^{\prime \prime}$ is the imaginary component of the
magnetic susceptibility. Recall that $\gamma$ is constant for a freely 
rotating grain with
dynamic symmetry. Rather than attempt a detailed analysis of the torque
for the case of a grain lacking dynamic symmetry, we simply adopt the 
Davis-Greenstein result in equation (\ref{eq:DG-torque}), replacing
$\sin^2 \gamma$ with $\langle \sin^2 \gamma \rangle$. 

\section{Simulations} \label{sec:simulations}

In this section, we describe the fundamental elements of the simulation
codes used in this work. 

\subsection{Dimensionless variables}

For numerical integration of the equations of motion, we adopt dimensionless
variables
\be
J^{\prime} = \frac{J}{I_1 \omega_T}
\ee
and
\be
t^{\prime} = \frac{t}{\tau_{\mathrm{drag}}} ,
\ee
where the thermal rotation rate is
\begin{multline}
\omega_T = \left( \frac{15 k T_{\mathrm{gas}}}{8 \upi \bar{\rho} a_{\mathrm{eff}}^5}
\right)^{1/2} \\
= 1.6573 \times 10^5 \left( \frac{\bar{\rho}}{3 \, \mathrm{g} \, 
\mathrm{cm}^{-3}} \right)^{-1/2} \left( \frac{T_{\mathrm{gas}}}{100 \, 
\mathrm{K}} \right)^{1/2} \left( \frac{a_{\mathrm{eff}}}{0.1 \, \mu 
\mathrm{m}} \right)^{-5/2} \mathrm{s}^{-1}
\label{eq:omega-T}
\end{multline}
and the drag time-scale is 
\begin{multline}
\tau_{\mathrm{drag}} = \frac{3 I_1 \delta^{4/3}}{4 \sqrt{\upi} m n v_{\mathrm{th}}
a_{\mathrm{eff}}^4} = 1.045 \times 10^5 \, \alpha_1 \delta^{4/3} \\
\times \left(
\frac{\bar{\rho}}{3 \, \mathrm{g} \, \mathrm{cm}^{-3}} \right) \left( 
\frac{a_{\mathrm{eff}}}{0.1 \, \mu \mathrm{m}} \right) \left( \frac{n}
{30 \, \mathrm{cm}^{-3}} \right)^{-1} \left( \frac{T_{\mathrm{gas}}}{100 \,
\mathrm{K}} \right)^{-1/2} \mathrm{yr} .
\label{eq:tau-drag}
\end{multline}

\subsection{Internal relaxation} \label{subsec:int-relax}

In dimensionless variables, the Langevin equation for internal relaxation 
becomes
\be
\label{eq:internal-Langevin-eq-dimensionless}
\mathrm{d}q = - A_1(J^{\prime}, q) \, 
\frac{\tau_{\mathrm{drag}}}{\tau_{\mathrm{int}}(J^{\prime})}
\, \mathrm{d}t^{\prime} + B_1[b(J^{\prime}), q] \,
\sqrt{\frac{\tau_{\mathrm{drag}}}{\tau_{\mathrm{int}}(J^{\prime})}} \, 
\mathrm{d}w^{\prime}_{\mathrm{int}} 
\ee
where $\mathrm{d}w^{\prime}_{\mathrm{int}}$ is a Gaussian random variable with
variance $\mathrm{d}t^{\prime}$, 
\be
A_1(J^{\prime}, q) = - \tau_{\mathrm{int}}(J^{\prime}) \, A(q) ,
\ee
and 
\be
B_1[b(J^{\prime}), q] = \sqrt{\tau_{\mathrm{int}}(J^{\prime}) \, D[b(J^{\prime}), q]} .
\ee
Random numbers and Gaussian random variables are computed using modified
versions of the routines {\sc ran2} and {\sc gasdev} from \citet{Press92}.
The internal drift and diffusion coefficients $A_1(q)$ and $B_1(b,q)$
are computed using equations (55), (61), and (81) in \citet{KW17}, adopting
their assumption that $D[b(J^{\prime}), q] = 0$ when $q=r_3$. The function
$b(J^{\prime})$ is defined by 
\be
\label{eq:b-def}
b(J^{\prime}) = \frac{J^2}{2 I_1 k T_d} = 
\frac{\alpha_1}{2} \, \frac{T_{\mathrm{gas}}}{T_{\mathrm{d}}}(J^{\prime})^2 .
\ee
From the final two paragraphs in Section \ref{sec:nuclear}, we take
\be
\tau_{\mathrm{int}}(J^{\prime}) = \left\{ \left[ \tau_{\mathrm{Bar}}(J^{\prime})
\right]^{-1} + \left[ \tau_{\mathrm{nuc}}(J^{\prime}) \right]^{-1} \right\}^{-1} ,
\label{eq:tau-int-defined}
\ee
\begin{multline}
\tau_{\mathrm{Bar}}(J^{\prime}) = 6.77 \times 10^{6} \, \alpha_1 \left( 
\frac{\bar{\rho}}{3 \, \mathrm{g} \, \mathrm{cm}^{-3}} \right)^2 \left(
\frac{a_{\mathrm{eff}}}{0.1 \, \mu \mathrm{m}} \right)^7 \left( 
\frac{T_{\mathrm{gas}}}{100 \, \mathrm{K}} \right)^{-1} \\
\times \left( 
\frac{T_{\mathrm{d}}}{15 \, \mathrm{K}} \right) \left( J^{\prime} \right)^{-2}
\, \mathrm{s} \ ,
\end{multline}
\begin{multline}
\tau_{\mathrm{nuc}}(J^{\prime}) = 1.36 \times 10^{-5} \, \tau_{\mathrm{Bar}}(J^{\prime})
\bigg\{ 1 + 410 \bigg[ \left( \frac{\bar{\rho}}{3 \, \mathrm{g} \, 
\mathrm{cm}^{-3}} \right)^{-1/2} \\
\times \left( \frac{T_{\mathrm{gas}}}{100 \, \mathrm{K}} 
\right)^{1/2} \left( \frac{a_{\mathrm{eff}}}{0.1 \, \mu \mathrm{m}} \right)^{-5/2} 
\bigg]^{1.96} \left( J^{\prime} \right)^{1.96} \bigg\}^{1.02} .
\label{eq:tau-nuc-defined}
\end{multline}

\begin{figure}
\includegraphics[width=90mm]{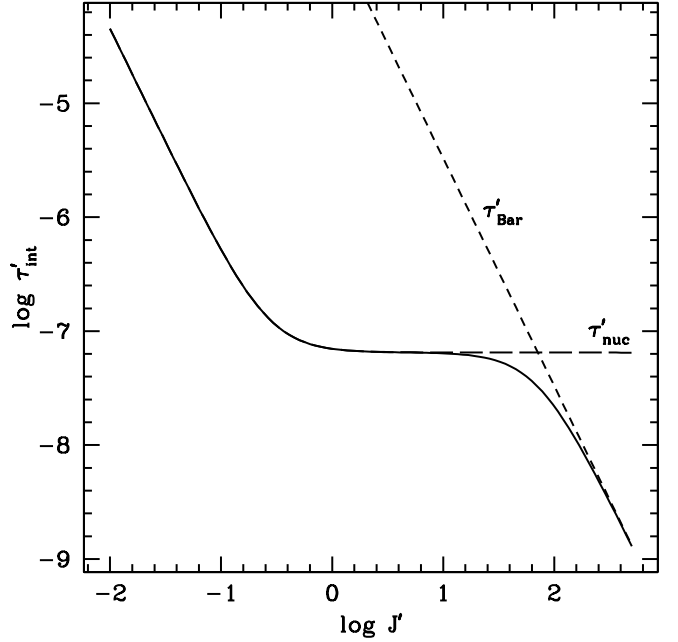}
\caption{
Internal relaxation times normalized to the drag time, from equations
(\ref{eq:tau-int-defined})--(\ref{eq:tau-nuc-defined}), for
$r_2 = 1.3$, $r_3 = 1.5$, $\bar{\rho} = 3 \, \mathrm{g} \, \mathrm{cm}^{-3}$,
$T_{\mathrm{gas}} = 100 \, \mathrm{K}$, $T_d = 15 \, \mathrm{K}$, and
$a_{\mathrm{eff}} = 0.2 \, \mu \mathrm{m}$.
        }
\label{fig:int-relax-times}
\end{figure}

Fig.~\ref{fig:int-relax-times} shows the internal relaxation times
from the above three equations versus $J^{\prime}$, adopting parameter values
suitable for a silicate grain in the cold neutral medium (CNM):  
$r_2 = 1.3$, $r_3 = 1.5$, $\bar{\rho} = 3 \, \mathrm{g} \, \mathrm{cm}^{-3}$,
$T_{\mathrm{gas}} = 100 \, \mathrm{K}$, $T_d = 15 \, \mathrm{K}$, and
$a_{\mathrm{eff}} = 0.2 \, \mu \mathrm{m}$. 
(Throughout this paper,  we will denote the base-10 logarithm by `log' and the 
natural logarithm by `ln'.) When following the evolution of
$q$ in simulations, we must ensure that the time step size 
$\mathrm{d}t^{\prime} \ll \tau^{\prime}_{\mathrm{int}}$. 

We construct interpolation tables for $A_1(q)$ and $B_1(b, q)$, using
{\sc mathematica}. As seen in
Fig. 1 in \citet{KW17}, these functions approach zero very steeply at
$q=r_2$. They also approach zero at $q=r_3$, as does $A(q)$ at $q=1$. 
In order to obtain precise values of $A_1(q)$ and $B_1(b, q)$ for all $q$,
we construct tables for $B_1(b,q)$ for six separate ranges
of $q$: (1) $q \in [1 + 10^{-15}, 1.01]$ with uniform spacing in 
$\ln(q-1)$, (2) $q \in [1.01, 1.28]$ with uniform spacing in 
$q$, (3) $q \in [1.28, 1.3 - 10^{-15}]$ (recall that $r_2=1.3$) with 
uniform spacing in $\ln(r_2-q)$, (4) $q \in [1.3 + 10^{-15}, 1.32]$
with uniform spacing in $\ln(q-r_2)$, (5) $q \in [1.32, 1.49]$ with 
uniform spacing in $q$, (6)  $q \in [1.49, 1.5 - 10^{-15}]$ (recall 
that $r_3 = 1.5$) with uniform spacing in $\ln(r_3-q)$.
For each range, we take 2000 values of $q$. We 
take 451 values of $b$, spaced uniformly in $\ln b$, from 
$b_{\mathrm{min}} = 4.45 \times 10^{-10}$ to $b_{\mathrm{max}}= 5 \times 10^5$. 
As seen in equation (81) in \citet{KW17}, $B_1(b, q)$ becomes independent of
$b$ as $b \rightarrow 0$. Thus, for $b < b_{\mathrm{min}}$, the minimum value of
$b$ in the interpolation tables, we simply take $b = b_{\mathrm{min}}$. Over the
full range of $q$, the fractional error due to this approximation is always
$< 10^{-10}$. 

We also construct interpolation tables for $A_1(q)$ in ranges (2)--(5). 
For $A_1(q)$ in ranges (1) and (6), we employ the asymptotic formulas in
\citet{KW17}, their equations (59) and (62). These are modified slightly
as $q$ deviates from 1 and $r_3$ to ensure that the asymptotic formulas
yield exactly the same result as the interpolation table when $q = 1.01$
and $q = 1.49$, which are the (1)--(2) and (5)--(6) boundaries, respectively.
Thus, we take
\be
A_1(q) = c_{A<} \, (q-1) \left[ 1 - \upsilon_< (q-1) \right] \ \ \ , 
\ \ \ q < 1.01
\ee
and
\be
A_1(q) = c_{A>} \, (r_3 - q) \left[ 1 - \upsilon_> (r_3 - q) \right] \ \ \ , 
\ \ \ q > 1.49
\ee
with 
\be
c_{A<} = \frac{r_3^2 (r_2-1) + r_2^2 (r_3 - 1)}{2} ,
\ee
\be
c_{A>} = \frac{r_3 - r_2 + r_2^2 (r_3 - 1)}{2} ,
\ee
\be
\upsilon_< = 100 \left[ 1 - \frac{100 A_1(q = 1.01)}{c_{A<}}  \right] ,
\ee
and
\be
\upsilon_> = 100 \left[ 1 - \frac{100 A_1(q = 1.49)}{c_{A>}}  \right] .
\ee
These expressions are exact when $q=1.01$ and $1.49$ and have fractional 
errors less than $3 \times 10^{-6}$ in range (1) and $2 \times 10^{-5}$ in 
range (6).

Lacking a first-principles theory of Barnett relaxation, it is not clear
how to treat the boundaries at $q=1$ and $q=r_3$. In test runs, we 
found frequent overshooting of $q=1$ (where the diffusion coefficient is
non-zero) but not of $q=r_3$ (where the diffusion coefficient is taken to
be zero). In test simulations assuming thermal equilibrium, we found that the
distribution function for $q$ resulting from the simulation best agrees with
the theoretical distribution if a `reflecting' boundary condition is adopted
at $q=1$. That is, if the value of $q$ resulting from the Langevin equation
is $1-\epsilon$, then we instead set $q=1+\epsilon$. 
The treatment of the boundary at $q=r_3$ does not significantly
affect the results. Thus, we adopt `reflecting' boundary conditions at both
$q=1$ and $q=r_3$.
In contrast, 
\citet{W09} and \citet{KW17} reduced the time step
and repeated the step whenever the simulation overshot $q=1$. Since this
prescription involves discarding randomly chosen variables,
it can introduce statistical biases. Thus, we reject that prescription
here.


\subsection{Angular momentum evolution}

In dimensionless variables, the Langevin equations for the grain's 
angular-momentum components become
\be
\label{eq:external-Langevin-eq-dimensionless}
\mathrm{d}J^{\prime}_{i, J} = \langle \Gamma^{\prime}_{i, J}(\mathbfit{J}^{\prime}, q, 
\mathrm{fs}) \rangle \, \mathrm{d}t^{\prime} + \sum_{j=1}^3 \langle 
B^{\prime}_{ij, J}(\mathbfit{J}^{\prime}, q, \mathrm{fs}) \rangle
\, \mathrm{d}w^{\prime}_{j, J} \ \ \ \ (i=1-3) ,
\ee
where $\mathrm{d}w^{\prime}_{j, J}$ are Gaussian random variables with 
variance $\mathrm{d}t^{\prime}$, 
\be
\label{eq:Gamma-prime}
\langle \Gamma^{\prime}_{i, J}(\mathbfit{J}^{\prime}, q, \mathrm{fs}) \rangle = 
\frac{\tau_{\mathrm{drag}} \langle \Gamma_{i, J}(\mathbfit{J}^{\prime}, q, 
\mathrm{fs}) \rangle}{I_1 \omega_T} , 
\ee
\be
\label{eq:B-prime}
\langle B^{\prime}_{ij, J}(\mathbfit{J}^{\prime}, q, \mathrm{fs}) \rangle = 
\frac{\tau^{1/2}_{\mathrm{drag}} \langle B_{ij, J}(\mathbfit{J}^{\prime}, q, 
\mathrm{fs}) \rangle}{I_1 \omega_T} ,
\ee
and 
$\langle \Gamma_{i, J}(\mathbfit{J}^{\prime}, q, \mathrm{fs}) \rangle$ and
$\langle B_{ij, J}(\mathbfit{J}^{\prime}, q, \mathrm{fs}) \rangle$ are the 
components of the rotationally averaged mean torque and diffusion tensor
resulting from all of the external processes under consideration,
respectively.

In the following sections, we will assume that all of the particles departing
the grain surface do so either via thermal evaporation or within H$_2$ 
molecules formed at special surface sites. 

\subsection{Collisions and thermal evaporation}

First consider the case that all of the particles departing the grain surface 
do so via thermal evaporation. 
From equations (\ref{eq:Gamma-col-avg}), (\ref{eq:Gamma-ev-avg}),
(\ref{eq:avg-quantity}), (\ref{eq:app-i1}), (\ref{eq:app-i2}), 
(\ref{eq:app-i4}), and (\ref{eq:Gamma-prime}), the dimensionless rotationally 
averaged mean torque, arising from both collisions and evaporation, is given by
\begin{multline}
\langle \bmath{\Gamma^{\prime}} \rangle_{\mathrm{col + ev}}  = \langle 
\bmath{\Gamma^{\prime}}_{\mathrm{col}} \rangle + \langle 
\bmath{\Gamma^{\prime}}_{\mathrm{ev}} \rangle \\
= - \left[ Z_1(\delta) 
\langle \cos^2 \gamma \rangle + Z_2(\delta) \left( q - \langle \cos^2 \gamma 
\rangle \right) \right] \bmath{J^{\prime}} ,
\label{eq:mean-torque-thermal-eq}
\end{multline}
where $Z_1(\delta)$ and $Z_2(\delta)$ are defined in equations (\ref{eq:Z-1}) 
and (\ref{eq:Z-2}), respectively. 
Note that $Z_1(\delta) \rightarrow 1$ and $Z_2(\delta) \rightarrow 1$
as $\delta \rightarrow 1$ (i.e.~as the spheroid approaches a sphere).
Thus, $\langle \bmath{\Gamma^{\prime}} \rangle  = - \mathbfit{J}^{\prime}$
for a sphere, motivating the definition of the drag time-scale in equation
(\ref{eq:tau-drag}). 

From equations (\ref{eq:C-zz-J-avg}), (\ref{eq:C-xx-J-avg}), 
(\ref{eq:C-ij-ev}), and (\ref{eq:B-prime}), 
the dimensionless rotationally averaged diffusion tensor, arising from both 
collisions and evaporation, is given by
\be
\label{eq:Czz-thermal-eq}
\langle C^{\prime}_{zz, J} \rangle_{\mathrm{col + ev}} 
= \alpha_1^{-1} \left( 1 + \frac{T_{\mathrm{ev}}}
{T_{\mathrm{gas}}} \right) \left[ Z_1(\delta) \langle \cos^2 \gamma
\rangle + Z_2(\delta) \langle \sin^2 \gamma \rangle \right] ,
\ee
\begin{multline}
\langle C^{\prime}_{xx, J} \rangle_{\mathrm{col + ev}} = \langle C^{\prime}_{yy, J} 
\rangle_{\mathrm{col + ev}} \\
= \alpha_1^{-1} \left( 1 + \frac{T_{\mathrm{ev}}}{T_{\mathrm{gas}}} \right) \ 
\frac{1}{2} 
\left[ Z_2(\delta) \left( 1 + \langle \cos^2 \gamma \rangle \right) 
+ Z_1(\delta) \langle \sin^2 \gamma \rangle \right] .
\label{eq:Cxx-thermal-eq}
\end{multline}
For thermal equilibrium, $T_{\mathrm{ev}} = T_{\mathrm{gas}}$. 

\citet{RL99} derived the mean torque and diffusion coefficients due to
collisions and evaporation for an oblate spheroid with dynamic symmetry. They
presented results in an inertial frame, corresponding to our alignment
coordinates. With $T_{\mathrm{ev}} = T_d$ and 
$r_2=r_3$, our expressions (equations 
\ref{eq:mean-torque-thermal-eq}--\ref{eq:Cxx-thermal-eq}, with the diffusion
tensor transformed from angular-momentum to alignment coordinates using 
equations \ref{eq:x-J}--\ref{eq:z-J}) reduce to
to the \citet{RL99} expressions (their equations A10--A17). Note that
our $Z_1(\delta)$ and $Z_2(\delta)$ correspond to their functions
$\Gamma_{\parallel}(e)$ and $\Gamma_{\perp}(e)$, respectively. Note also that
they define their dimensionless quantities somewhat differently than we do,
so that the ratio of our dimensionless mean torque to theirs equals 
$Z_1(\delta)$ and the ratio of our dimensionless diffusion tensor components
to theirs equals $Z_1(\delta)/\alpha_1$. 

\subsection{Collisions and H$_2$ formation at special sites}
\label{sec:coll-H2-special}

Now consider the case that all of the particles departing the grain surface 
do so within H$_2$ molecules formed at special surface sites. 
From equations (\ref{eq:Gamma-col-avg}), (\ref{eq:Gamma-H2-sys-rot-avg}),
and (\ref{eq:Gamma-H2-drag}), the dimensionless systematic and drag torques are
\begin{multline}
\langle \bmath{\Gamma^{\prime}}_{\mathrm{H2, sys}} \rangle = \frac{3 v_{\mathrm{H2}}}
{4 a_{\mathrm{eff}} \omega_T} \delta^{1/3} \left[ 1 + \delta^2 g(\delta) \right] \\
\times 
\left( Q_1 \langle \cos \gamma \rangle + Q_2 \langle \sin \gamma \cos \alpha
\rangle \right) \bmath{\hat{J}} ,
\label{eq:H2-sys-dimensionless}
\end{multline}
\begin{multline}
\langle \bmath{\Gamma^{\prime}}_{\mathrm{drag}} \rangle_{\mathrm{col + H2}} = 
\langle \bmath{\Gamma^{\prime}}_{\mathrm{col}} \rangle +
\langle \bmath{\Gamma^{\prime}}_{\mathrm{H2, drag}} \rangle = \\
- \frac{3}{8} J^{\prime} \bmath{\hat{J}} \left[ \left( 1 - \delta^2 \right)^2 
\mathcal{I}_4(\delta) q + \left\{ 2 \left[ 1 + \delta^2 g(\delta) \right]
Q_3 - \left( 1 - \delta^2 \right)^2 \mathcal{I}_4(\delta) \right\} \right. \\
\times \langle \cos^2 \gamma \rangle 
+ 2 \left[ 1 + \delta^2 g(\delta) \right] \Big( Q_4
\langle \sin^2 \gamma \sin^2 \alpha \rangle \\
+ Q_5 \langle \sin^2 \gamma
\cos^2 \alpha \rangle \Big) \bigg] .
\label{eq:torque-drag-plus-H2}
\end{multline}
From equations (\ref{eq:C-xx-J-H2-sites}) and (\ref{eq:C-zz-J-H2-sites}),
only the diagonal elements of $C_{ij, \mathrm{H}2}$ are needed. From equation
(\ref{eq:C-ij-H2-special-sites}), 
\be
C^{\prime}_{xx, \mathrm{H}2} = \frac{3 m v^2_{\mathrm{H2}}}{2 k_B T_{\mathrm{gas}}} \
\alpha_1^{-1} \left[ 1 + \delta^2 g(\delta) \right] Q_6 \, ;
\label{eq:C-xx-prime-H2}
\ee
$C^{\prime}_{yy, \mathrm{H}2}$ and $C^{\prime}_{zz, \mathrm{H}2}$ are of identical form,
with $Q_6$ replaced by $Q_7$ and $Q_8$, respectively. 
The components of the dimensionless rotationally averaged diffusion tensor for
collisions are given in equations (\ref{eq:Czz-thermal-eq}) and
(\ref{eq:Cxx-thermal-eq}), 
omitting the term $T_{\mathrm{ev}} / T_{\mathrm{gas}}$.

\subsection{Davis-Greenstein torque}

From equation (\ref{eq:DG-torque}), and the last sentence in Section
\ref{sec:D-G}, we take the dimensionless Davis-Greenstein torque to be
\be
\label{eq:DG-torque-dimensionless}
\langle \bmath{\Gamma}^{\prime}_{\mathrm{DG}} \rangle = - \frac{\tau_{\mathrm{drag}}}
{\tau_{\mathrm{DG}}} \left[ 1 + \left( r_2 - 1 \right) \langle \sin^2 \gamma 
\rangle \right]
\left( J^{\prime}_{x, B} \, \bmath{\hat{x}}_B + J^{\prime}_{y,B}
\bmath{\hat{y}}_B \right) .
\ee
It is somewhat more efficient to compute $\mathrm{d}\mathbfit{J}$ due to the
Davis-Greenstein torque in alignment coordinates rather than in 
angular-momentum coordinates.

\subsection{Omission of the Barnett torque}

The term 
$J^{\prime}_{x, B} \, \bmath{\hat{x}}_B + J^{\prime}_{y,B} \bmath{\hat{y}}_B$
in equation (\ref{eq:DG-torque-dimensionless}) can also be expressed as
$J^{\prime} \sin \xi ( \bmath{\hat{\xi}} \cos \xi + \bmath{\hat{J}} \sin \xi)$.
As seen in equations 
(\ref{eq:mean-torque-thermal-eq})--(\ref{eq:C-xx-prime-H2})
and the preceding sentence, 
none of the torques or diffusion coefficients considered so far depend
explicitly on the coordinate $\phi_B$. The
Barnett torque, due to the interaction of the grain's Barnett magnetic moment 
with the interstellar magnetic field, yields a rapid precession of 
the grain angular momentum about the field direction, i.e.~rapid change of
$\phi_B$. Since none of the processes here depend on $\phi_B$, for simplicity
we omit the Barnett torque and set $\phi_B = 0$ at the end of each time step.

\subsection{Rotational averages}

As described in Section \ref{sec:external-processes} and seen in equations
(\ref{eq:mean-torque-thermal-eq})--(\ref{eq:DG-torque-dimensionless}),  
several rotational 
averages involving the Eulerian angles (e.g.~$\langle \cos^2 \gamma \rangle$)
are needed. As derived in Section \ref{sec:grain-rotation}, these are
expressed in terms of $\langle \mathrm{dn}^2(\nu, k^2) \rangle$,
$\langle \mathrm{cn}^2(\nu, k^2) \rangle$, and $\upi/[2 K(k^2)]$. We tabulate
each of these quantities for $10^4$ values of $k^2$ between 0 and 0.9999
and interpolate. When $k^2 > 0.9999$, we employ the approximations
\be
\langle \mathrm{dn}^2(\nu, k^2) \rangle \approx \left[ \ln \left(
\frac{4}{\sqrt{1 - k^2}} \right) \right]^{-1}
\ee
and
\be
K(k^2) \approx \left( 1 + \frac{1-k^2}{4} \right) \ln \left(
\frac{4}{\sqrt{1 - k^2}} \right) - \frac{1-k^2}{4} ;
\ee
$\langle \mathrm{cn}^2(\nu, k^2) \rangle$ is obtained from identity
(\ref{eq:cn2-avg}).

\subsection{Thermal averages over $q$}

\begin{figure}
\includegraphics[width=90mm]{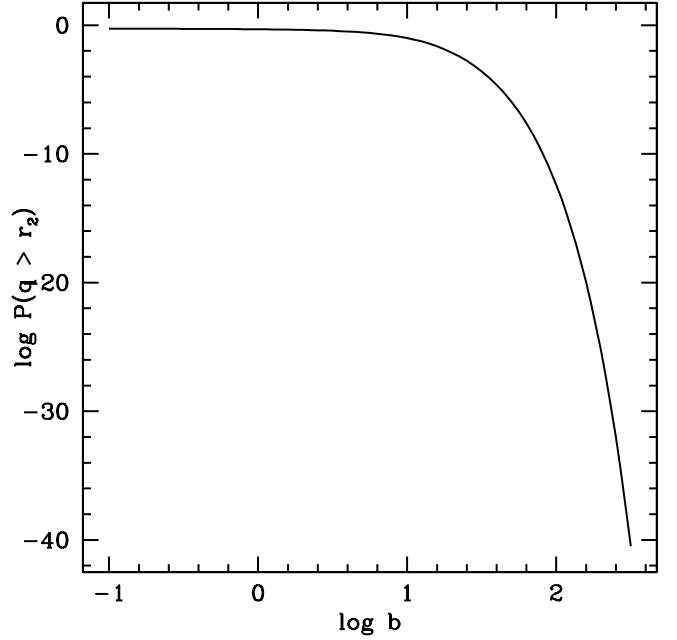}
\caption{
The probability that $q > r_2$ for a thermal distribution, as a function of
$b$.
        }
\label{fig:prob-q-r2}
\end{figure}

For sufficiently low $J^{\prime}$, $q$ can reach values higher than $r_2$
and flipping can occur. Of course, it is necessary to follow the evolution
of $q$ in this case. For high $J^{\prime}$, the flipping probability is 
negligible, so it is sufficient to average the torque and diffusion 
coefficients, assuming a thermal distribution of $q$ values. 
Fig.~\ref{fig:prob-q-r2} shows the probability that $q>r_2$ for a thermal
distribution, as a function of $b$. We follow the evolution of $q$ when 
$b < b_{\mathrm{crit}} = 300$ and average over $q$ when $b > b_{\mathrm{crit}}$,
using equation (63) in \citet{KW17} for the thermal-equilibrium distribution
of $q$. Interpolation tables for thermally averaged quantities were generated
using {\sc mathematica}. Whenever $b$ crosses $b_{\mathrm{crit}}$ from above,
a value of $q$ is randomly chosen from its thermal-equilibrium distribution
for $b = b_{\mathrm{crit}}$. 

\subsection{Time step size} \label{sec:time-step-size}

In the high-$b$ regime, for which we average over $q$, we take a constant
time step size $\mathrm{d}t^{\prime} = k_{\mathrm{high}}$.  In the low-$b$ regime, 
we take the step size 
$\mathrm{d}t^{\prime} = k_{\mathrm{low}} \tau_{\mathrm{int}}^{\prime}$. 
In most simulations, we take $k_{\mathrm{low}} = 10^{-2}$ and 
$k_{\mathrm{high}} = 10^{-4}$.

\subsection{Grain flipping}

When $q < r_2$ ($q > r_2$), the grain is in either the positive or negative
flip state with respect to $\bmath{\hat{a}}_1$ ($\bmath{\hat{a}}_3$).
Whenever $q=r_2$ is crossed, the flip state is chosen randomly, with equal
probability to be positive or negative. A grain flips when it starts with
$q<r_2$ in one flip state with respect to $\bmath{\hat{a}}_1$, makes an
excursion to $q > r_2$, and returns to $q<r_2$ in the opposite flip state
with respect to $\bmath{\hat{a}}_1$.

\section{Code test: thermal equilibrium}

\label{sec:code-test}

In order to test both the theoretical development in Section 
\ref{sec:external-processes} and our simulation code, we performed simulations
with two simplified codes adopting thermal equilibrium. 

\subsection{Distribution of $q$}

\begin{figure}
\includegraphics[width=90mm]{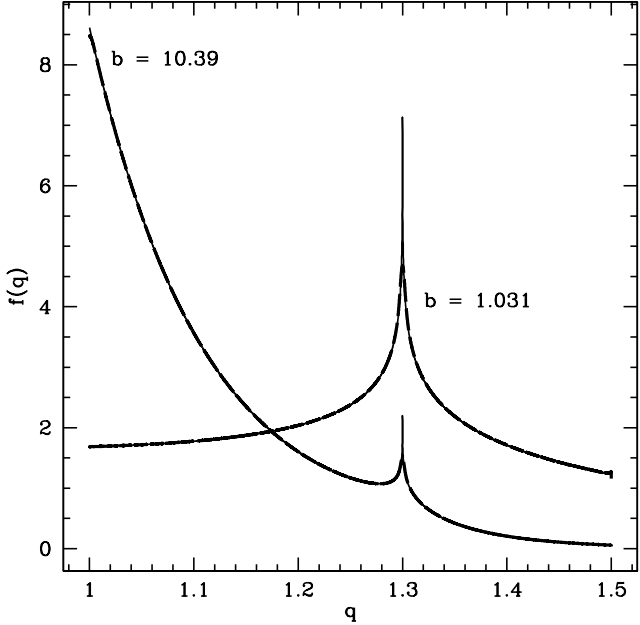}
\caption{
The thermal-equilibrium
distribution function $f(q)$ from simulations (heavy dashed curves) and
theory (light solid curves) for two values of $b$, as indicated.
        }
\label{fig:q-hist-eq}
\end{figure}

In the first code, external processes are omitted; the angular 
momentum and grain temperature are held constant and $q$ is evolved using
equation (\ref{eq:internal-Langevin-eq-dimensionless}). In these 
simulations, the time step is taken to be $10^{-3} \, \tau_{\mathrm{int}}$ and
the the total duration of the simulation is $10^8 \, \tau_{\mathrm{int}}$. 
The code returns the distribution function $f(q)$; the fraction of the time
that the grain has dimensionless energy parameter between $q$ and 
$q+\mathrm{d}q$ is $f(q) \mathrm{d}q$. These are compared with the 
theoretical distribution $f_{\mathrm{TE}}(q)$, given by equation (63) in 
\citet{KW17}. Fig.~\ref{fig:q-hist-eq} shows results from the simulations
(heavy dashed curves) and theory (light solid curves) for two values of 
$b$ (defined in eq.~\ref{eq:b-def}), 
$b=1.031$ and $b=10.39$. There is a vertical asympotote at $q=r_2=1.3$,
which the analytic equation better captures (with a finite number of values
of $q$ for which $f_{\mathrm{TE}}$ is evaluated) than the simulation. Otherwise,
apart from slight deviations at $q=1$ and $q=r_3$, the results are identical.

\subsection{Distribution of $J^{\prime}$}

In the second test, we check that the simulation correctly reproduces the 
distribution $f(J^{\prime})$, defined such that the fraction of the time that
the grain has dimensionless angular momentum between $J^{\prime}$ and
$J^{\prime} + \mathrm{d} J^{\prime}$ equals $f(J^{\prime}) \, \mathrm{d} J^{\prime}$,
in thermal equilibrium. 

\subsubsection{Simulation} \label{sec:thermal-equil-J-dist-sim}

In this code, the only
external processes are collisions and thermal evaporation. Thus, the mean
torque and diffusion coefficients are given by equations 
(\ref{eq:mean-torque-thermal-eq})--(\ref{eq:Cxx-thermal-eq}), with 
$T_{\mathrm{ev}} = T_{\mathrm{gas}} = T_d$. The actual value of the temperature does 
not matter, since we are examining the distribution of $J^{\prime}$ rather than
the distribution of $J$. In order to obtain results in a reasonable run time, 
we average all quantities over a thermal distribution of $q$, for all values 
of $b$. We take the time step and duration of the simulation equal to 
$10^{-5} \, \tau_{\mathrm{drag}}$ and $10^5 \, \tau_{\mathrm{drag}}$, respectively.
Table \ref{tab:parameters} indicates the adopted parameter values.

\begin{table*}
\caption{Adopted parameter values for simulations.} 
\label{tab:parameters}
\begin{tabular}{llllll}
\hline
symbol &
quantity & 
section \ref{sec:thermal-equil-J-dist-sim} &
section \ref{sec:bias} & 
section \ref{sec:crossovers} & 
section \ref{sec:D-G-alignment} \\
& & thermal equil & f-step bias & crossovers suite 1 & D-G case 1 \\
\hline
$\delta$ & spheroid semilength ratio & 0.5 & 0.5 & 0.5 & 0.5 \\
$r_2$ & $I_1/I_2$ & 1.3 & 1.3 & 1.3 & 1.3 \\
$r_3$ & $I_1/I_3$ & 1.5 & 1.5 & 1.5 & 1.5 \\
$\bar{\rho}$ & grain mean density (g$\,$cm$^{-3}$) & 3.0 & 3.0 & 3.0 & 3.0 \\
$n_{\mathrm{H}}$ & gas H number density (cm$^{-3}$) & 30 & 30 & 30 & 30 \\
$a_{\mathrm{eff}}$ & grain effective radius ($\mu$m) & 0.2 & 0.2 & 0.2 & 0.2 \\
$T_{\mathrm{gas}}$ & gas temperature (K) & NA & 100 & 100 & 100 \\
$T_d$ & dust temperature (K) & NA & 15 & 15 & 15 \\
$N_s$ & number of H$_2$-formation sites & NA & $5.5 \times 10^5$ & 
$5.5 \times 10^5$ & $5.5 \times 10^5$ \\
$E_{\mathrm{H}2}$ & H$_2$ kinetic energy (eV) & NA & 0.2 & 0.2 & 0.2 \\
$(\cos \theta_{\mathrm{out}})_{\mathrm{min}}$ & see Section \ref{sec:H2-formation} & 
NA & 0.8 & 0.8 & 0.8 \\
$B$ & interstellar magnetic field ($\mu$G) & NA & NA & NA & 5.0 \\
$t^{\prime}_{\mathrm{life}}$ & H$_2$-formation site lifetime & NA & NA & NA & 1 \\
$k_{\mathrm{low}}$ & low-$b$ time-step parameter & NA & $10^{-2}$ & $10^{-2}$ & 
$10^{-2}$ \\
$k_{\mathrm{high}}$ & high-$b$ time-step parameter & $10^{-5}$ & NA & $10^{-4}$ & 
$10^{-4}$ \\
$t^{\prime}$ & duration of simulation (if fixed) & $10^5$ & NA & NA & $10^3$ \\ 
\hline
\end{tabular}
\end{table*}


\subsubsection{Theoretical distribution function}

\label{sec:theoretical-dist-funcs}

For a freely rotating body, the Lagrangian $L$ equals the rotational kinetic
energy $E$, 
\be
\label{eq:Lagrangian}
L = E = \frac{1}{2} I_1 \omega_1^2 + \frac{1}{2} I_2 \omega_2^2 + 
\frac{1}{2} I_3 \omega_3^2
\ee
and the square of the angular momentum is
\be
J^2 = I_1^2 \omega_1^2 + I_2^2 \omega_2^2 + I_3^2 \omega_3^2 .
\ee

The components of the angular velocity along the principal axes can be
expressed in terms of the Eulerian angles:
\be
\omega_1 = \dot{\zeta} \cos \gamma + \dot{\alpha} ,
\ee
\be
\omega_2 = \dot{\zeta} \sin \alpha \sin \gamma + \dot{\gamma} \cos \alpha ,
\ee
\be
\omega_3 = \dot{\zeta} \cos \alpha \sin \gamma - \dot{\gamma} \sin \alpha ,
\ee
where dots denote time derivatives. 
Inserting these into equation (\ref{eq:Lagrangian}) for the Lagrangian, 
the momenta conjugate to the Eulerian angles are easily obtained; 
e.g.~$p_{\zeta} = \partial L / \partial \dot{\zeta}$. After some algebra, we
find $E$ and $J$ in terms of the Eulerian angles and their conjugate 
momenta, yielding
\begin{multline}
J^{\prime} = \left\{ \left( p^{\prime}_{\alpha} \right)^2 + \left[ \cos \alpha \,
p^{\prime}_{\gamma} + \frac{\sin \alpha}{\sin \gamma} \left( p^{\prime}_{\zeta} - 
\cos \gamma \, p^{\prime}_{\alpha} \right) \right]^2 + \right. \\
\left. \left[ \sin \alpha \, p^{\prime}_{\gamma}
- \frac{\cos \alpha}{\sin \gamma} \left( p^{\prime}_{\zeta} - \cos \gamma \,
p^{\prime}_{\alpha} \right) \right]^2 \right\}^{1/2} ,
\end{multline}
\begin{multline}
q = \left( J^{\prime} \right)^{-2}
\left\{ \left( p^{\prime}_{\alpha} \right)^2 + r_2 \left[ \cos \alpha \,
p^{\prime}_{\gamma} + \frac{\sin \alpha}{\sin \gamma} \left( p^{\prime}_{\zeta} - 
\cos \gamma \, p^{\prime}_{\alpha} \right) \right]^2 \right. \\
\left. + r_3 \left[ \sin \alpha \,
p^{\prime}_{\gamma}
- \frac{\cos \alpha}{\sin \gamma} \left( p^{\prime}_{\zeta} - \cos \gamma \,
p^{\prime}_{\alpha} \right) \right]^2 \right\} ,
\end{multline}
where $p_i^{\prime} = p_i / (I_1 \omega_T)$. 

The grain rotational states are uniformly distributed in the 
6-dimensional phase space defined by the Eulerian angles and their 
conjugate momenta, but neither $J^{\prime}$ nor $q$ depends explicitly on
$\zeta$. 
Define the density of states $\rho(J^{\prime}, q)$ such that the number of
states with dimensionless angular momentum and rotational enegy between
$J^{\prime}$ and $J^{\prime} + \mathrm{d}J^{\prime}$ and $q$ and $q+\mathrm{d}q$ is
proportional to
$\rho(J^{\prime}, q) \, \mathrm{d}J^{\prime} \, \mathrm{d}q$.
To estimate the density of states, we calculate $J^{\prime}$
and $q$ for $(230)^5$ combinations of 
$(\gamma, \alpha, p^{\prime}_{\zeta}, p^{\prime}_{\gamma}, p^{\prime}_{\alpha})$,
with the angles uniformly distributed from $0$ to $2 \upi$ and the conjugate
momenta distributed logarithmically from $7 \times 10^{-4}$ to $3$. We have
tried other values for the number of combinations and range of values of the
momenta and found that the distribution function for the relevant range of
$J^{\prime}$ and $q$ is well converged. 

The thermal-equilibrium distribution of $J^{\prime}$ is 
\be
f(J^{\prime}) = 
\int_1^{r_3} \mathrm{d}q \, \rho(J^{\prime}, q) \, \exp \left[ - \frac{\alpha_1}{2}
\left( J^{\prime} \right)^2 \right] ;
\ee
the form of the Boltzmann factor follows from equations (\ref{eq:q}), 
(\ref{eq:I_i}), and (\ref{eq:omega-T}). The temperature does not explicitly 
appear because the variable is $J^{\prime}$ rather than $J$. 

\subsubsection{Results}

\begin{figure}
\includegraphics[width=90mm]{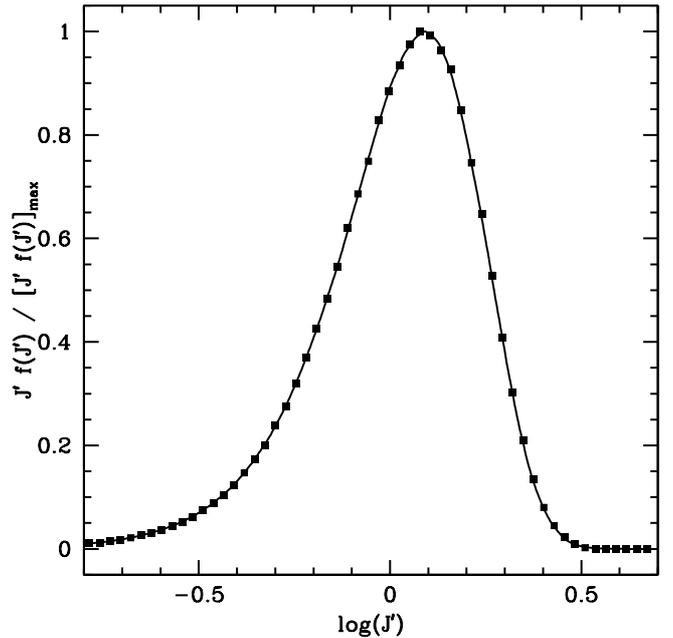}
\caption{
$J^{\prime} f(J^{\prime})$, normalized to its maximum value, 
for thermal equilibrium. The solid curve is for
the simulation results and the boxes are the theoretical result. 
        }
\label{fig:J-hist-eq}
\end{figure}

Fig.~\ref{fig:J-hist-eq} shows the results from both simulation and theory,
which agree perfectly.

\section{H$_2$ formation} \label{sec:H2-formation}

In the following sections, we will assume that all H atoms depart the grain
surface within H$_2$ molecules, formed at special surface sites.

\citet{DW97} took the surface area per H$_2$-formation site equal
to $l^2$, with $l^2 = 100 \,$\AA$^2$ as a fiducial value. They also 
considered larger values of $l^2$. The surface area of an oblate spheroid is
\be
S = 2 \upi a_{\mathrm{eff}}^2 \delta^{-2/3} \left[ 1 + \delta^2 g(\delta) \right] .
\ee
For $a_{\mathrm{eff}} = 0.2 \, \mu \mathrm{m}$ and $\delta = 0.5$, 
$S = 5.5 \times 10^7 \,$\AA$^2$. Thus, for 
$l^2 = 100 \,$\AA$^2$, the number of H$_2$-formation sites is
$N_s = 5.5 \times 10^5$. We will adopt this value in our simulations.

The translational kinetic energy of the ejected H$_2$ molecule is not well
known. In most simulations, we take $E_{\mathrm{H}2} = 0.2 \, \mathrm{eV}$, 
but we also perform some simulations with $E_{\mathrm{H}2} = 0.05 \, \mathrm{eV}$ 
and $1.0 \, \mathrm{eV}$. 

At the start of each simulation, the locations of all $N_s$ sites, as well
as the departure directions at the sites, are chosen randomly as described
in Section \ref{sec:special-sites}, taking
$(\cos \theta_{\mathrm{out}})_{\mathrm{min}} = 0.8$.
To provide a sense for the magnitude of the dimensionless efficiency factors 
$Q_i$ ($i$ = 1--8) defined in Section \ref{sec:special-sites}, Table
\ref{tab:special-sites} tabulates their values for 
one particular grain realization with $N_s = 5.5 \times 10^5$.

\begin{table}
\caption{Efficiency factors associated with H$_2$ formation for one grain
realization with number of special sites $N_s = 5.5 \times 10^5$.} 
\label{tab:special-sites}
\begin{tabular}{ll}
\hline
$Q_1$ &   $-4.00 \times 10^{-4}$ \\
$Q_2$ &   $3.13 \times 10^{-4}$ \\
$Q_3$ &   $0.600$ \\     
$Q_4$ &   $0.521$ \\    
$Q_5$ &   $0.599$ \\    
$Q_6$ &   $9.04 \times 10^{-2}$ \\    
$Q_7$ &   $9.01 \times 10^{-2}$ \\    
$Q_8$ &   $5.40 \times 10^{-2}$ \\
\hline
\end{tabular}
\end{table}

If a grain reaches suprathermal rotation, then $q \approx 1$ and 
$\gamma \approx 0$. Setting the systematic and drag torques 
(equations \ref{eq:H2-sys-dimensionless} and \ref{eq:torque-drag-plus-H2})
equal yields the equilibrium value of the dimensionless angular momentum:
\be
J^{\prime}_{\mathrm{eq}} = \frac{\delta^{1/3} Q_1 v_{\mathrm{H2}}}{Q_3 a_{\mathrm{eff}}
\omega_T} .
\label{eq:Jp_eq_from_Qs}
\ee
In this expression, a positive (negative) value corresponds to rotation with
positive (negative) flip state with respect to $\bmath{\hat{a}}_1$.
Given the relevant parameters in Table \ref{tab:parameters} and the efficiency
factors in Table \ref{tab:special-sites},
$J^{\prime}_{\mathrm{eq}} = 395$. In equilibrium, this grain is in
the negative flip state with respect to $\bmath{\hat{a}}_1$.

\begin{figure}
\includegraphics[width=90mm]{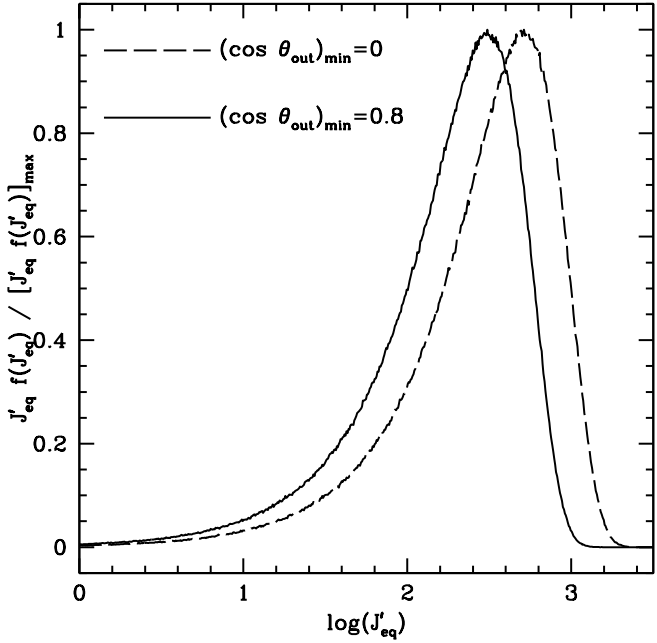}
\caption{
Distribution of $J^{\prime}_{\mathrm{eq}}$ for randomly constructed grain
realizations with $(\cos \theta_{\mathrm{out}})_{\mathrm{min}} = 0$ or 0.8
(see text). Parameters are as in Table \ref{tab:parameters}, with 
$E_{\mathrm{H2}} = 0.2 \, \mathrm{eV}$, and $N_s = 5.5 \times 10^5$. The
distribution function $f(J^{\prime}_{\mathrm{eq}})$ is defined such that the
fraction of grain with dimensionless equilibrium angular momentum between
$J^{\prime}_{\mathrm{eq}}$ and
$J^{\prime}_{\mathrm{eq}} + \mathrm{d}J^{\prime}_{\mathrm{eq}}$ equals 
$f(J^{\prime}_{\mathrm{eq}}) \, \mathrm{d}J^{\prime}_{\mathrm{eq}}$.
        }
\label{fig:Jp-eq-hist}
\end{figure}

Fig.~\ref{fig:Jp-eq-hist} shows the distribution of $J^{\prime}_{\mathrm{eq}}$
derived from $6.4 \times 10^6$  different grain realizations randomly 
constructed as described in Section \ref{sec:special-sites}, with the
relevant parameters as in Table \ref{tab:parameters}, except for   
$(\cos \theta_{\mathrm{out}})_{\mathrm{min}}$ equal to both 0 and 0.8. 

From equations (\ref{eq:C_zz-body}), (\ref{eq:C_xx-body}), 
(\ref{eq:C-ij-H2-special-sites}), and (\ref{eq:Q_6-def})--(\ref{eq:Q_8-def}),
in the limit of uniform surface coverage of formation sites with 
$(\cos \theta_{\mathrm{out}})_{\mathrm{min}} = 0$, $Q_6$ and 
$Q_7 \rightarrow Q_{6, \mathrm{un}}$ and $Q_8 \rightarrow Q_{8, \mathrm{un}}$, where
\be
Q_{6, \mathrm{un}} = \frac{Z_2(\delta)}{3 [1 + \delta^2 g(\delta)]} ,
\ee
\be
Q_{8, \mathrm{un}} = \frac{Z_1(\delta)}{3 [1 + \delta^2 g(\delta)]} .
\ee
Recall that the functions $Z_1(\delta)$ and $Z_2(\delta)$ are defined in
equations (\ref{eq:Z-1}) and (\ref{eq:Z-2}). For $\delta = 0.5$,
$Q_{6, \mathrm{un}} = 0.1187$ and $Q_{8, \mathrm{un}} = 0.1501$. 
In our suite of $6.4 \times 10^6$ different grain realizations with 
$(\cos \theta_{\mathrm{out}})_{\mathrm{min}} = 0$, $Q_6$ and $Q_7$ are always
within 1\% of $Q_{6, \mathrm{un}}$ and $Q_8$ is always within 1\% of 
$Q_{8, \mathrm{un}}$. For the suite with 
$(\cos \theta_{\mathrm{out}})_{\mathrm{min}} = 0.8$, $Q_6$ and $Q_7$ are always
close to $0.76 \, Q_{6, \mathrm{un}}$ and $Q_8$ is always close to 
$0.36 \, Q_{8, \mathrm{un}}$.

In Section \ref{sec:crossovers}, where we examine the duration of and 
disalignment during a single crossover, the efficiency factors 
$Q_1$--$Q_8$ associated with H$_2$ formation are held fixed throughout the
simulation. In Section \ref{sec:D-G-alignment}, examining Davis-Greenstein
alignment, existing formation sites are destroyed and new formation sites
are formed, so that $Q_1$--$Q_8$ vary throughout the simulation. For 
simplicity, and to keep the total number of special formation sites, 
$N_s$, constant, we assume that when a site is destroyed, another
site forms at the same time. The probability that no sites are destroyed
during time $\Delta t_{\mathrm{rep}}$ is given by 
\be
P_{\mathrm{none}} = \exp \left( - \frac{N_s \Delta t_{\mathrm{rep}}}{t_{\mathrm{life}}}
\right) ,
\ee
where $t_{\mathrm{life}}$ is the site lifetime. Thus, we take the time until
the next site-replacement event to be 
\be
\Delta t_{\mathrm{rep}} = - \frac{t_{\mathrm{life}} \ln P}{N_s} ,
\ee
where $P$ is a random number between 0 and 1. During a site-replacement
event, one existing site is randomly chosen for destruction. A new site
is immediately formed, with parameters chosen randomly following the same
prescription for constructing the sites at the start of the simulation.
We perform simulations with three values of 
$t^{\prime}_{\mathrm{life}} = t_{\mathrm{life}}/\tau_{\mathrm{drag}} = 1$, 10, and
$10^3$.

\section{F-step duration bias} \label{sec:bias}

In this section, we present an example to illustrate the f-step duration
bias described in Section \ref{sec:intro}. Recall that we define an f-step
as the interval between consecutive flips and define up-steps and down-steps
as f-steps with 
$\mathbfit{J} \bmath{\cdot} \Gamma_1 \bmath{\hat{a}}_1 > 0$ and
$\mathbfit{J} \bmath{\cdot} \Gamma_1 \bmath{\hat{a}}_1 < 0$, respectively.
In an up-step, the systematic torque acts so as to spin the grain up to
higher angular momentum; the opposite applies in a down-step.

We ran four suites of simulations, with parameter values as given in Table
\ref{tab:parameters}. The special sites are held fixed and
are identical for all of the simulations. The values of the efficiency
factors associated with H$_2$ formation are given in Table
\ref{tab:special-sites}. The Davis-Greenstein torque is not included. The
time step parameter $k_{\mathrm{low}} = 10^{-2}$; $k_{\mathrm{high}}$ is not 
relevant, since the high-$b$ regime never applies. 

Each suite consists of $4 \times 10^6$ simulations and each simulation starts
with $J^{\prime} = 2$ and $q = 1.01$. Each simulation is terminated when the
flip state with respect to $\bmath{\hat{a}}_1$ changes. Thus, each simulation
corresponds approximately to one f-step. Note that
\citet{KW17} adopted a more realistic,
but also more complicated, definition of a flip; the simple approximation
here is sufficient for our purpose. Suites 1 and 2 are for an up-step
and a down-step, respectively. Likewise for suites 3 and 4,
except that the parameter $b$, which depends on $J^{\prime}$ and is defined in
equation (\ref{eq:b-def}), is held fixed throughout the simulation in these
cases. To be clear, the value of $J^{\prime}$ changes in all of the simulations,
but in suites 3 and 4, we do not adjust the value of $b$ accordingly. Only
the internal relaxation is affected by the value of $b$; by holding it constant,
we artificially eliminate the dependence of the flipping probability per unit
time on $J^{\prime}$. 

Table \ref{tab:bias} shows the mean value of the (dimensionless) f-step
duration, $t^{\prime}_{\mathrm{av}}$, and the mean value of the change in the
(dimensionless) angular momentum, $(\Delta J^{\prime})_{\mathrm{av}}$, for each
suite. Both $t^{\prime}_{\mathrm{av}}$ and $|(\Delta J^{\prime})_{\mathrm{av}}|$ are
nearly identical for suites 3 and 4, as expected since the internal relaxation 
is taken to be independent of $J^{\prime}$ in these simulations. Also as 
expected, the 
values of $t^{\prime}_{\mathrm{av}}$ and $(\Delta J^{\prime})_{\mathrm{av}}$ are larger
for suite 1 (up-steps) than for suites 3 and 4, and are both smaller for suite
2 (down-steps). 

\begin{table}
\caption{Mean f-step duration and change in angular momentum.}
\label{tab:bias}
\begin{tabular}{llll}
\hline
Suite & Description & $t^{\prime}_{\mathrm{av}}$ & $(\Delta J^{\prime})_{\mathrm{av}}$ \\
\hline
1 & up-step & $2.54 \times 10^{-5}$ & $5.78 \times 10^{-3}$ \\ 
2 & down-step & $2.39 \times 10^{-5}$ & $-5.44 \times 10^{-3}$ \\ 
3 & up-step (fixed $b$) & $2.46 \times 10^{-5}$ & $5.59 \times 10^{-3}$ \\ 
4 & down-step (fixed $b$) & $2.46 \times 10^{-5}$ & $-5.60 \times 10^{-3}$ \\ 
\hline
\end{tabular}
\end{table}

\begin{figure}
\includegraphics[width=90mm]{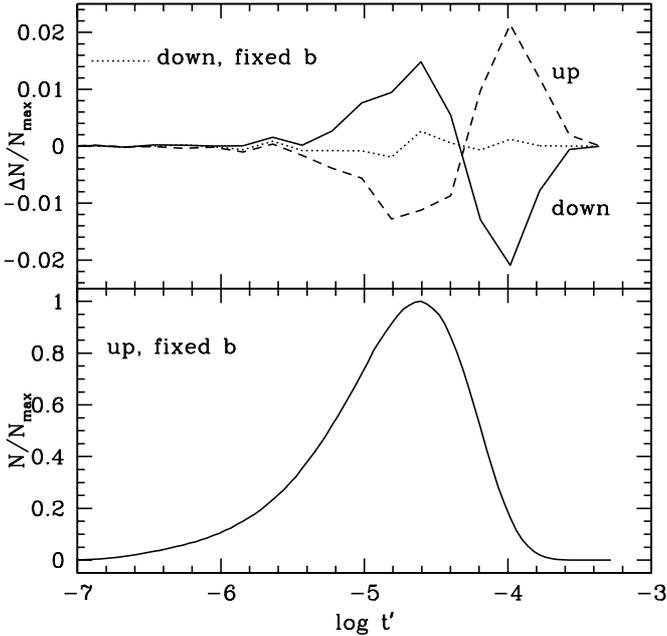}
\caption{
Bottom panel: Histogram $N/N_{\mathrm{max}}$ of $\log t^{\prime}$ for suite 3 
(up-steps, constant $b$), normalized at its peak. Top panel: The difference 
$\Delta N/N_{\mathrm{max}}$ between the histogram for suites 1 (up-steps, dashed),
2 (down-steps, solid), and 4 (down-steps, fixed $b$, dotted). 
        }
\label{fig:bias}
\end{figure}

The bottom panel of Fig.~\ref{fig:bias} shows the histogram $N/N_{\mathrm{max}}$
of $\log t^{\prime}$ for suite 3, normalized at its peak. The top panel shows
the difference $\Delta N/N_{\mathrm{max}}$ between the histogram for each of the 
other three suites and suite 3. The difference is relatively small for suite 4;
it should approach zero as the number of simulations per suite increases.
For suite 1 (up-steps), there is an excess at the longest times, compensated
by a deficit at shorter times. The opposite trend applies for suite 2 
(down-steps).

\section{Crossovers} \label{sec:crossovers}

In this section, we examine the duration of individual crossovers, as well
as the disalignment during the crossover. Each simulation begins with the
construction of the special sites, as described in Section 
\ref{sec:H2-formation}. We demand that $J^{\prime}_{\mathrm{eq}} \ge 50$;
otherwise, the construction is discarded and a new one is generated.
The initial dimensionless angular momentum is set equal to 
$J^{\prime}_c  = (1 - e^{-1}) J^{\prime}_{\mathrm{eq}} \approx 0.632 \, J^{\prime}_{\mathrm{eq}}$. The flip state is chosen such that the systematic torque spins the
grain down, to lower values of $J^{\prime}$. During the simulation, the special
sites are held fixed. The simulation ends when 
$J^{\prime} = J^{\prime}_c$, in the flip state such that the systematic torque
is spinning the grain up. 

These simulations include internal relaxation and only two external 
processes: collisions and H$_2$ formation at special sites (section
\ref{sec:coll-H2-special}). The Davis-Greenstein torque is not included. 
Initially, the grain is oriented such that the alignment angle
$\xi = 0$. Thus, the value of $\cos \xi$ at the end of the simulation 
indicates the disalignment: $\cos \xi = 1$ implies no angular deviation of
the angular momentum vector during the crossover. 

Now suppose that the grain is constrained to always rotate about
$\bmath{\hat{a}}_1$ and is only subject to the mean systematic and drag
torques. That is, stochastic elements are neglected. In this case, the
grain spins down from $J^{\prime}_c$ to $J^{\prime} = 0$ and then back up to
$J^{\prime}_c$, without any flip and with the final angular momentum pointing
in the opposite direction as the initial angular momentum. In this case,
the duration $t^{\prime}_s$ of the crossover (normalized to the drag time)
can be found analytically:
\be
t^{\prime}_s = \frac{4}{3 [1 + \delta^2 g(\delta)] Q_3} \ln \left( \frac{1 + 
J^{\prime}_c/J^{\prime}_{\mathrm{eq}}}{1 - J^{\prime}_c/J^{\prime}_{\mathrm{eq}}} \right)
\approx \frac{1.987}{[1 + \delta^2 g(\delta)] Q_3} .
\ee
If thermal trapping is important, then we would expect that 
$t^{\prime}/t^{\prime}_s \gg 1$, where $t^{\prime}$ is the actual duration of the
crossover as found in the simulation.

We ran four suites of simulations. The parameter values for suite 1 
are given in Table \ref{tab:parameters}. Each suite consists of 
$5.6 \times 10^4$ separate simulations with identical input parameter values;
only the seed for the random number generator differs among the simulations
within a suite. For suites 2 and 3, different values are adopted for the 
kinetic energy of the outgoing H$_2$ molecules:
$E_{\mathrm{H}2} = 0.05 \, \mathrm{eV}$ and
$1.0 \, \mathrm{eV}$. For suites 1--3, the step-size parameters
(see Section \ref{sec:time-step-size}) are taken to be $k_{\mathrm{low}} = 10^{-2}$
and $k_{\mathrm{high}} = 10^{-4}$. Suite 4 serves as a convergence
check, with $E_{\mathrm{H}2} = 0.2 \, \mathrm{eV}$, $k_{\mathrm{low}} = 10^{-3}$,
and $k_{\mathrm{high}} = 10^{-5}$.

\begin{figure}
\includegraphics[width=90mm]{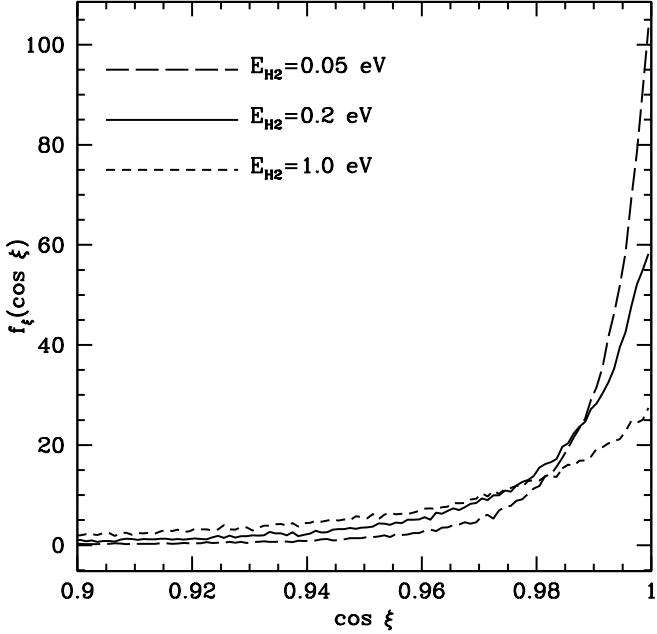}
\caption{
The distribution function $f_{\xi}(\cos \xi)$ for suites of crossover simulations
with the energy $E_{\mathrm{H}2}$ of the departing H$_2$ molecule as 
indicated. See the text for other parameter values. 
        }
\label{fig:cross-hist-cxi}
\end{figure}

\begin{figure}
\includegraphics[width=90mm]{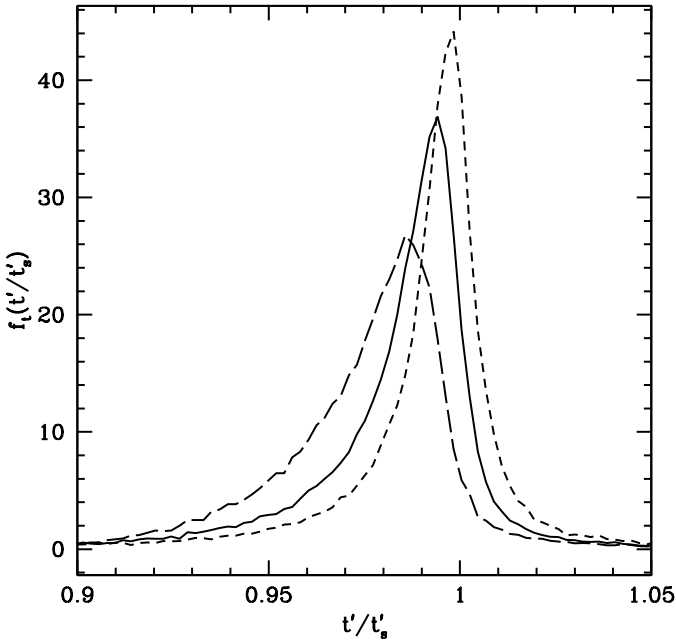}
\caption{
Same as Figure \ref{fig:cross-hist-cxi} except for the distribution function
$f_t(t^{\prime}/t^{\prime}_s)$. Line types as in Figure \ref{fig:cross-hist-cxi}.
        }
\label{fig:cross-hist-rat}
\end{figure}

\begin{figure}
\includegraphics[width=90mm]{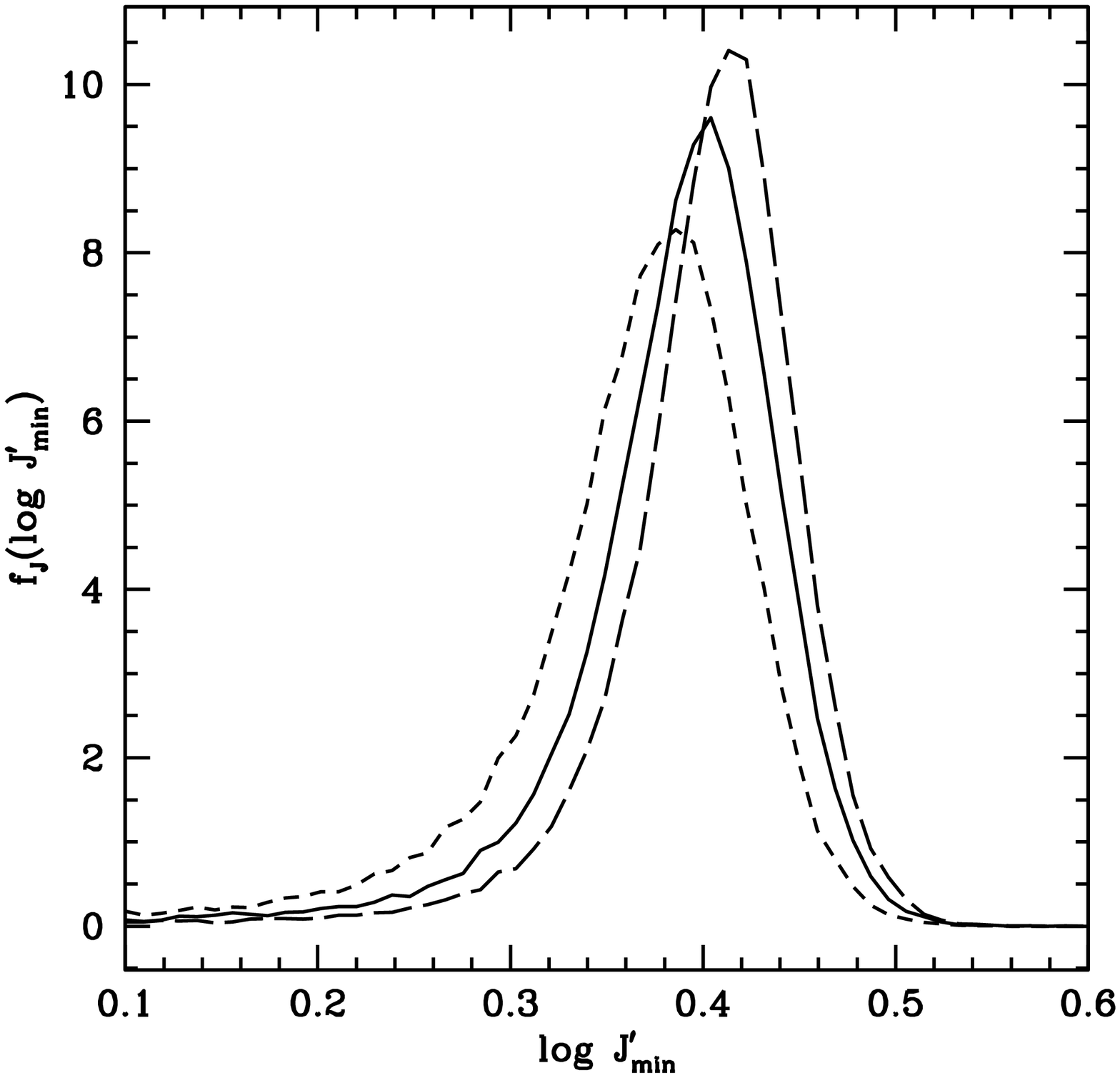}
\caption{
Same as Figure \ref{fig:cross-hist-cxi} except for the distribution function
$f_J(\log J^{\prime}_{\mathrm{min}})$. Line types as in Figure 
\ref{fig:cross-hist-cxi}.
        }
\label{fig:cross-hist-Jp-min}
\end{figure}

Figures \ref{fig:cross-hist-cxi}--\ref{fig:cross-hist-Jp-min} show the
following three distribution functions for suites 1--3:
(1) $\cos \xi$; (2) $t^{\prime}/t^{\prime}_s$, the ratio
of the crossover duration $t^{\prime}$ to the duration $t^{\prime}_s$ for the
simple case described above; and (3) $\log J^{\prime}_{\mathrm{min}}$, where
$J^{\prime}_{\min}$ is the minimum value of $J^{\prime}$ during the crossover.
In each case, $f_i(u) \, \mathrm{d}u$ equals the fraction of simulations for
which the argument lies between $u$ and $u+\mathrm{d}u$. The three distribution
functions are denoted by subscripts `$\xi$', `$t$', and `$J$', respectively,
for $\cos \xi$, $t^{\prime}/t^{\prime}_s$, and $\log J^{\prime}_{\mathrm{min}}$.

Generally, the disalignment is mild and the duration is comparable to
$t^{\prime}_s$. That is, thermal trapping is not a prevalent condition. 
The distribution
functions display clear trends as a function of the H$_2$ kinetic energy.
As $E_{\mathrm{H}2}$ increases, the distributions shift towards more 
disalignment, longer duration, and lower minimum value of $J^{\prime}$. 
Table \ref{tab:outliers} shows the fraction of the simulations within each
suite for which $\cos \xi$ and $J^{\prime}_{\mathrm{min}}$ are less than the
lower limits in Figs.~\ref{fig:cross-hist-cxi} and
\ref{fig:cross-hist-Jp-min} and for which $t^{\prime}/t^{\prime}_s$ is greater
than the upper limit in Fig.~\ref{fig:cross-hist-rat}. The tails of the
distributions favor longer durations and lower $J^{\prime}_{\mathrm{min}}$ for
lower $E_{\mathrm{H}2}$, though the precision far out in the tails is, of 
course, low. 


\begin{table}
\caption{Fraction of the crossover simulations satisfying outlying 
conditions.}
\label{tab:outliers}
\begin{tabular}{llll}
\hline
Condition & $E_{\mathrm{H}2} = 0.05 \, \mathrm{eV}$ &
$E_{\mathrm{H}2} = 0.2 \, \mathrm{eV}$ &
$E_{\mathrm{H}2} = 1.0 \, \mathrm{eV}$ \\
\hline
$\cos \xi < 0.9$ &  $3.55 \times 10^{-2}$ & 
$6.24 \times 10^{-2}$ & $0.195$ \\
$\cos \xi < 0.8$ &  $2.66 \times 10^{-2}$ & 
$2.67 \times 10^{-2}$ & $8.93 \times 10^{-2}$ \\
$\cos \xi < 0.5$ &  $1.79 \times 10^{-2}$ & 
$1.10 \times 10^{-2}$ & $2.35 \times 10^{-2}$ \\     
$\cos \xi < 0$ &  $1.07 \times 10^{-2}$ & 
$5.32 \times 10^{-3}$ & $6.62 \times 10^{-3}$ \\    
$\cos \xi < -0.5$ &  $4.80 \times 10^{-3}$ & 
$2.00 \times 10^{-3}$ & $2.43 \times 10^{-3}$ \\    
$t^{\prime}/t^{\prime}_s > 1.05$ &  $4.91 \times 10^{-2}$ & 
$3.30 \times 10^{-2}$ & $1.78 \times 10^{-2}$ \\    
$t^{\prime}/t^{\prime}_s > 2.0$ &  $1.98 \times 10^{-2}$ & 
$7.68 \times 10^{-4}$ & $0$ \\    
$t^{\prime}/t^{\prime}_s > 3.0$ &  $1.24 \times 10^{-2}$ & 
$5.36 \times 10^{-5}$ & $0$ \\    
$t^{\prime}/t^{\prime}_s > 4.0$ &  $7.73 \times 10^{-3}$ & 
$1.79 \times 10^{-5}$ & $0$ \\    
$t^{\prime}/t^{\prime}_s > 10.0$ &  $9.82 \times 10^{-4}$ & 
$0$ & $0$ \\    
$J^{\prime}_{\mathrm{min}} < 10^{-1}$ &  $9.88 \times 10^{-3}$ & 
$2.05 \times 10^{-3}$ & $1.34 \times 10^{-3}$ \\
$J^{\prime}_{\mathrm{min}} < 3 \times 10^{-2}$ & $5.07 \times 10^{-3}$ & 
$9.29 \times 10^{-4}$ & $5.89 \times 10^{-4}$ \\
$J^{\prime}_{\mathrm{min}} < 10^{-2}$ &  $3.11 \times 10^{-3}$ & 
$5.00 \times 10^{-4}$ & $2.86 \times 10^{-4}$ \\
$J^{\prime}_{\mathrm{min}} < 3 \times 10^{-3}$ & $1.89 \times 10^{-3}$ & 
$2.32 \times 10^{-4}$ & $7.14 \times 10^{-5}$ \\
$J^{\prime}_{\mathrm{min}} < 10^{-3}$ & $7.14 \times 10^{-4}$ & 
$3.57 \times 10^{-5}$ & $0$ \\
\hline
\end{tabular}
\end{table}

\begin{figure}
\includegraphics[width=90mm]{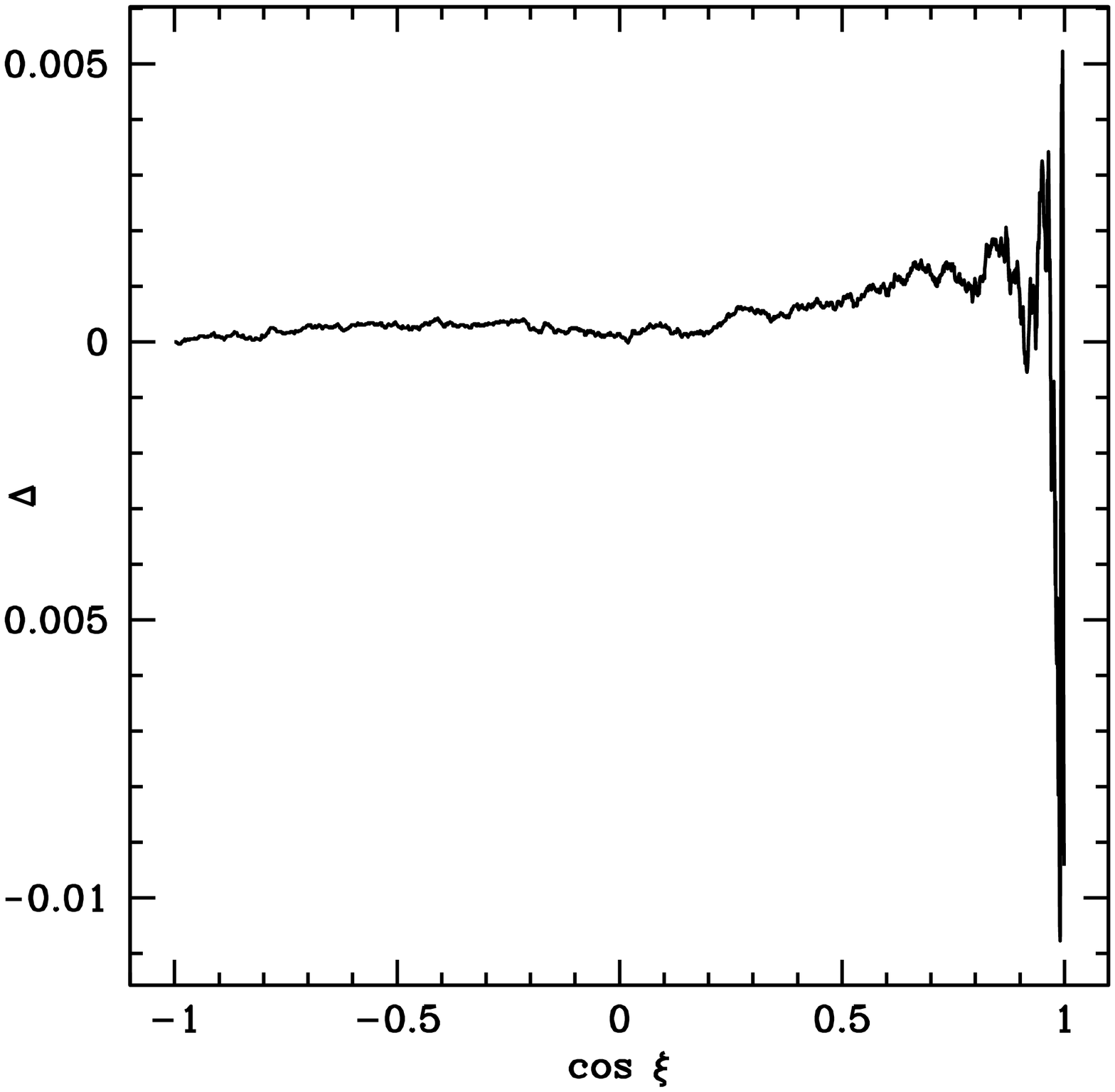}
\caption{
The fractional difference $\Delta(\cos \xi)$ between the cumulative 
distribution function (starting at $\cos \xi = 1$) for two simulation suites
with $E_{\mathrm{H}2} = 0.2 \, \mathrm{eV}$. The time-step parameters for the
two suites are $(k_{\mathrm{low}}, k_{\mathrm{high}}) = (10^{-2}, 10^{-4})$ and
$(10^{-3}, 10^{-5})$. 
        }
\label{fig:cross-conv}
\end{figure}

As noted above, simulation suites 1--3 take step-size parameters
$k_{\mathrm{low}} = 10^{-2}$ and $k_{\mathrm{high}} = 10^{-4}$. To check if these
yield sufficient convergence, suite 4 repeats suite 1, except with 
$k_{\mathrm{low}} = 10^{-3}$ and $k_{\mathrm{high}} = 10^{-5}$. We find that the
distribution functions are virtually identical for these two suites. 
The largest discrepancy is for $f_{\xi}(\cos \xi)$. To illustrate the
fine agreement even for this case, we examine
the cumulative distribution function,
$f_{\xi, \mathrm{cum}}(\cos \xi)$, starting at $\cos \xi = 1$. That is,
$f_{\xi, \mathrm{cum}}(\cos \xi) = \int_{\cos \xi}^1 f_{\xi}(u) \mathrm{d}u$.
Figure 
\ref{fig:cross-conv} shows $\Delta(\cos \xi)$ versus $\cos \xi$, where 
$\Delta(\cos \xi)$ is the fractional difference between the cumulative
distribution function for the two suites. 

As noted in Section \ref{subsec:int-relax}, we adopt `reflecting' boundary
conditions at $q=1$ and $q=r_3$, since this choice best reproduces the
distribution function for $q$ in the case of thermal equilibrium. Otherwise,
this choice is arbitrary. As a check on the extent to which this choice
affects the results, we repeated all four simulation suites with a different
prescription for the boundaries. In these runs, if a time step yields
$q < 1$ or $q > r_3$, then a new value of $\mathrm{d}w^{\prime}_{\mathrm{int}}$ is 
selected 
randomly and $\mathrm{d}q$ is recomputed. The distribution functions resulting 
from suites with the two different boundary prescriptions differ very slightly.

\section{Davis-Greenstein alignment} \label{sec:D-G-alignment}

In this section, we examine the efficiency of Davis-Greenstein alignment. 
We constructed seven suites of simulations which include internal relaxation,
collisions with gas atoms, H$_2$ formation at time-varying special sites
(Section \ref{sec:H2-formation}), and the Davis-Greenstein torque. Each 
suite consists of 504 simulations. At the start of each simulation, the 
grain is randomly constructed as in previous sections, the alignment angle
$\xi$ is randomly chosen from a uniform distribution in $\cos \xi$ 
(with $-1 \le \cos \xi \le 1$), and $J^{\prime}$ is set equal to 2.5, which
corresponds to the peak of the distribution $f_J(\log J^{\prime}_{\mathrm{min}})$
for the suite of crossover simulations with 
$E_{\mathrm{H}2} = 0.2 \, \mathrm{eV}$ (Fig.~\ref{fig:cross-hist-Jp-min}).

For our reference suite (`case 1'), we take parameter values as in Table
\ref{tab:parameters}; the total 
duration of the simulation (normalized to the drag time-scale)
$t^{\prime} = 10^3$. For each of the other suites, one or more of the
parameter values are changed, as indicated in Table \ref{tab:D-G-runs}.
In cases 2 and 3, the site lifetime is increased. In case 4, thermal trapping 
is artificially prohibited, in the following extreme manner: 
Each time the low-$b$ regime is entered, after the first $q=r_2$ crossing, the 
flip state (with respect to both $\bmath{\hat{a}}_1$ and $\bmath{\hat{a}}_3$)
is always chosen such as to spin the grain up. Case 5 is a convergence check,
with the two parameters $k_{\mathrm{low}}$ and $k_{\mathrm{high}}$ both reduced 
by an order of magnitude; the duration $t^{\prime}$ of the simulation is also
reduced by an order of magnitude to avoid unmanageable run times. In case 6,
the dust temperature is increased to $T_d = 20 \, \mathrm{K}$. Finally, in case
7, a smaller grain size, $a_{\mathrm{eff}} = 0.05 \, \mu \mathrm{m}$, 
is considered.
In this case, the number of special sites $N_s$ is reduced in proportion to
the grain surface area (i.e.~by a factor of 16). Since the internal relaxation
time $\tau_{\mathrm{int}}$ decreases for the smaller grain, the duration of the 
simulation is also decreased, to 100 drag times, to avoid unmanageable run
times. 

\begin{table*}
\caption{Cases for Davis-Greenstein simulations.}
\label{tab:D-G-runs}
\begin{tabular}{lllll}
\hline
Case & Difference from Case 1 & $\langle \mathrm{RRF_{av}}(10) \rangle$ & 
$\langle \mathrm{RRF_{av}}(10^2) \rangle$ & 
$\langle \mathrm{RRF_{av}}(10^3) \rangle$ \\
\hline
1 & NA & 0.038 & 0.187 & 0.265 \\
2 & $t^{\prime}_{\mathrm{life}} = 10^3$ & 0.046 & 0.269 & 0.698 \\
3 & $t^{\prime}_{\mathrm{life}} = 10$ & 0.014 & 0.190 & 0.413 \\
4 & prohibited thermal trapping (see text) & 0.028 & 0.201 & 0.413 \\
5 & $k_{\mathrm{low}}= 10^{-3}$, $k_{\mathrm{high}} = 10^{-5}$, $t^{\prime} =10^2$ &
0.039 & 0.152 & NA \\
6 & $T_d = 20 \, \mathrm{K}$ & 0.049 & 0.131 & 0.214 \\
7 & $a_{\mathrm{eff}} = 0.05 \, \mu \mathrm{m}$; $N_s = 34,400$; $t^{\prime} =10^2$ 
& 0.094 & 0.412 & NA \\
\hline
\end{tabular}
\end{table*}

\begin{figure}
\includegraphics[width=90mm]{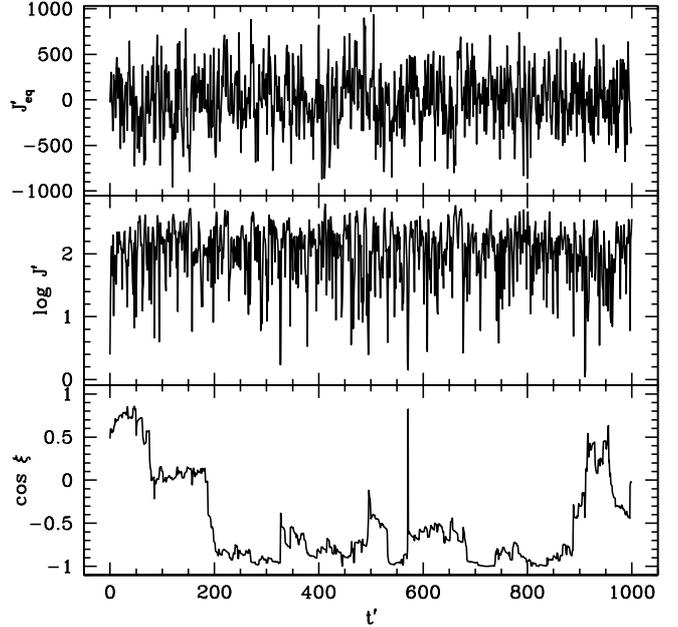}
\caption{
$J^{\prime}_{\mathrm{eq}}$, $\log J^{\prime}$, and $\cos \xi$ versus $t^{\prime}$
for one of the 504 case-1 Davis-Greenstein simulations. 
        }
\label{fig:example-case1}
\end{figure}

\begin{figure}
\includegraphics[width=90mm]{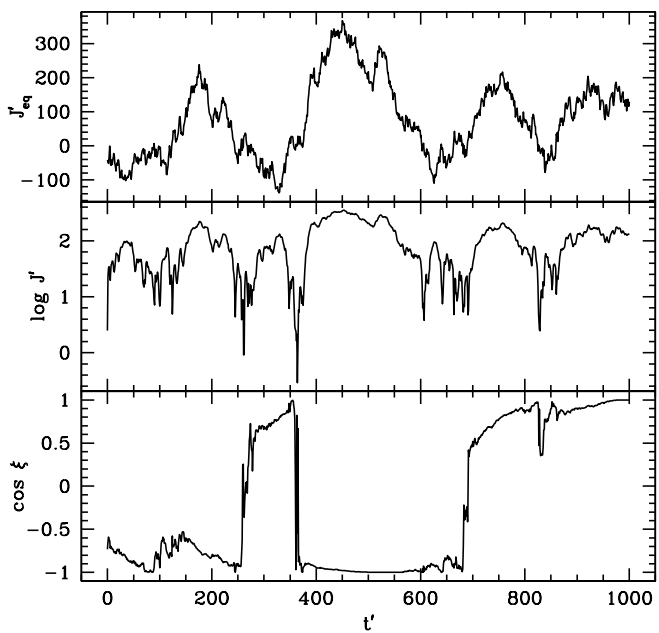}
\caption{
$J^{\prime}_{\mathrm{eq}}$, $\log J^{\prime}$, and $\cos \xi$ versus $t^{\prime}$
for one of the 504 case-2 Davis-Greenstein simulations. 
        }
\label{fig:example-case2}
\end{figure}

As an illustration of the grain dynamics, 
Figs.~\ref{fig:example-case1} and \ref{fig:example-case2} show 
$J^{\prime}_{\mathrm{eq}}$, $\log J^{\prime}$, and $\cos \xi$ versus $t^{\prime}$ for 
one of the 504 Davis-Greenstein simulations, for cases 1 and 2, 
respectively. These cases differ only in that $t^{\prime}_{\mathrm{life}} = 1$
for case 1 and $10^3$ for case 2. Consequently, $J^{\prime}_{\mathrm{eq}}$ and
$J^{\prime}$ fluctuate much more rapidly for case 1 than for case 2. The plots
are generated using $10^3$ output times (one per drag time); the fluctuations
of $J^{\prime}_{\mathrm{eq}}$ and $J^{\prime}$ are actually more pronounced than 
indicated in the figure, as would be seen if more output times were used in 
the figure construction. 

The Rayleigh reduction factor,
\be
\mathrm{RRF} = \frac{3}{2} \left( \cos^2 \xi - \frac{1}{3} \right) ,
\label{eq:RRF-defined}
\ee
is a useful measure of the grain alignment efficiency. 
In the Rayleigh limit, the linear dichroism of a grain rotating about 
$\bmath{\hat{a}}_1$ is proportional to RRF \citep{LD85}.
For alignment, RRF must exceed
zero; larger values of RRF correspond to higher degrees of alignment. 

\begin{figure}
\includegraphics[width=90mm]{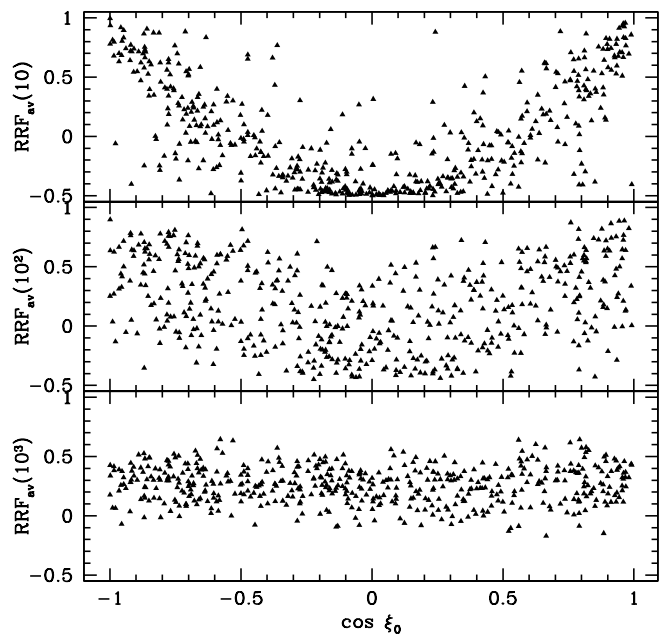}
\caption{
The Rayleigh reduction factor time-averaged over the first 10 drag times, the
first 100 drag times, and the full $10^3$-drag time simulation duration, 
versus $\cos \xi_0$, for the 504 case-1 Davis-Greenstein simulations. 
        }
\label{fig:RRF}
\end{figure}

Fig.~\ref{fig:RRF} shows $\mathrm{RRF_{av}}(10)$, 
$\mathrm{RRF_{av}}(10^2)$, and $\mathrm{RRF_{av}}(10^3)$ versus $\cos \xi_0$ for 
case 1, 
where $\mathrm{RRF_{av}}(t^{\prime})$ is the time-averaged Rayleigh reduction 
factor, from the start of the simulation to dimensionless time 
$t^{\prime} = t/\tau_{\mathrm{drag}}$, and $\xi_0$ is the initial value of $\xi$.
That is, the time average is taken over the
first 10 drag times, the first $10^2$ drag times, and the entire simulation 
(with duration $10^3$ drag times). Each point represents one of the 504
simulations. For the adopted parameters, 
$\tau_{\mathrm{DG}}/\tau_{\mathrm{drag}} = 73.3$. That is, the alignment time-scale
equals 73.3 times the drag time-scale. 
Thus, the plot of $\mathrm{RRF_{av}}(t^{\prime})$
versus $\cos \xi_0$ resembles that of equation (\ref{eq:RRF-defined}) for 
early $t^{\prime}$. By $t^{\prime} = 10^3$, the time-averaged Rayleigh reduction
factor no longer shows a dependence on the initial value of the alignment
angle $\xi_0$ and is positive for most simulations. 

For each of the seven simulation suites, table \ref{tab:D-G-runs} provides
the values of $\langle \mathrm{RRF_{av}}(t^{\prime}) \rangle$, the average of 
$\mathrm{RRF_{av}}(t^{\prime})$ over all 504 simulations (with $t^{\prime} = 10$,
$10^2$, and $10^3$, where applicable). On times 
$\sim 10^2$--$10^3 \, \tau_{\mathrm{drag}}$, paramagnetic dissipation does yield
partial grain alignment. As expected, the alignment is more efficient for
larger values of the H$_2$-formation site lifetime (cases 1--3). Thermal
trapping does not prevent D-G alignment, though the alignment may be somewhat
more efficient in case 4, where thermal trapping is artificially prohibited in 
an extreme manner. Comparing the results for cases 1 and 5, we conclude that
the adopted values for the time-step parameters $k_{\mathrm{low}}$ and 
$k_{\mathrm{high}}$ are reasonable. That is, the simulations appear to be
well converged. 

For all except case 7, the grain size
$a_{\mathrm{eff}} = 0.2 \, \mu \mathrm{m}$, for which (given the adopted 
CNM-like parameter values) $\tau_{\mathrm{drag}} = 1.3 \times 10^5 \, \mathrm{yr}$
and $\tau_{\mathrm{DG}}/\tau_{\mathrm{drag}} = 73$. Thus, times 
$\sim 10^2$--$10^3 \, \tau_{\mathrm{drag}}$ equate to 13--130$\,$Myr. 
From simulations of the multiphase interstellar medium, \citet{Peters17}
find a broad distribution of dust residence times in the CNM, with a median 
around $7 \,$Myr. Thus, the time required for D-G alignment of grains with
$a_{\mathrm{eff}} = 0.2 \, \mu \mathrm{m}$ seems uncomfortably long. 

Furthermore, $\tau_{\mathrm{drag}} \propto a_{\mathrm{eff}}$ and
$\tau_{\mathrm{DG}} \propto a_{\mathrm{eff}}^2$ (eqs.~\ref{eq:tau-drag} and
\ref{eq:tau-DG}). Thus, for $a_{\mathrm{eff}} = 0.05 \, \mu \mathrm{m}$, 
$\tau_{\mathrm{drag}} = 3.3 \times 10^4 \, \mathrm{yr}$ and 
$\tau_{\mathrm{DG}}/\tau_{\mathrm{drag}} = 18$. As seen from the case-7 result,
moderately efficient alignment is achieved on a time
$\sim 10^2 \, \tau_{\mathrm{drag}} \sim 3.3 \,$Myr. Thus, D-G alignment in the CNM
is more plausible for relatively small grains than for relatively large
grains, whereas observations reveal that only the relatively large grains are
well aligned \citep{KM95}. This contradiction between the model and 
observations has long plagued the D-G theory, even prior to the modifications
by Purcell. In their original work on thermal trapping, considering only 
Barnett relaxation, \citet{LD99a} found thermal trapping  
to be more severe for smaller grains, possibly resolving the contradiction.
However, with the introduction of nuclear relaxation, \citet{LD99b} concluded
that all grains are likely thermally trapped. We conclude that thermal 
trapping is not prevalent for either $a_{\mathrm{eff}} = 0.2 \, \mu \mathrm{m}$
or $a_{\mathrm{eff}} = 0.05 \, \mu \mathrm{m}$.

\citet{JS67} noted that if grains contain superparamagnetic 
inclusions, then the D-G alignment time-scale could be dramatically reduced.
If only the relatively large grains contain superparamagnetic inclusions,
then D-G alignment could be consistent with the \citet{KM95} results
\citep{M86}. We will examine D-G alignment for the case of grains with 
superparamagnetic inclusions in future work. From the results in this work,
we conclude that D-G alignment without superparamagnetic inclusions is
unlikely to account for alignment of the relatively large grains
responsible for the observed optical and infrared starlight polarization, in 
the diffuse ISM. However, it is the long 
alignment time, rather than any thermal trapping effect, that renders it
unlikely. 

\citet{HLM14} argued that small grains must be aligned to some extent
in order to explain the observed ultraviolet starlight polarization and
proposed that the observations could be used to estimate the interstellar
magnetic field strength. They examined D-G alignment of small grains, assuming 
that these grains are thermally trapped. Our results show that suprathermal
spin-up may be important for the small grains, potentially yielding higher
degrees of alignment. However, this conclusion is sensitive to the details of 
the H$_2$-formation model. We adopted $t^{\prime}_{\mathrm{life}} = 1$ in our
simulation with $a_{\mathrm{eff}} = 0.05 \, \mu \mathrm{m}$ (case 7), but smaller 
values are plausible and would yield a smaller degree of alignment. 
Furthermore, \citet{WD01force} argued that a model in which the grain surface
is saturated in chemisorption sites, as considered in section 
\ref{sec:uniform-sites} above, is plausible. In this case, any systematic 
torque would be negligible, precluding suprathermal spin-up. We will more
carefullly examine D-G alignment of small grains in future work. 

\section{Conclusions} \label{sec:conclusions}

In this study, we first extended the analysis of Barnett relaxation in 
\citet{KW17} beyond the low-frequency limit, enabling an approximate treatment 
of both Barnett and nuclear relaxation in thermally rotating grains. Since
no first-principles theory of Barnett or nuclear relaxation has been developed
to date, there is considerable uncertainty in the quantitative expressions
for the drift and diffusion coefficients. We followed \citet{KW17} in 
assuming that the diffusion coefficient goes to zero at $q=r_3$. We also 
neglected any deviation of the functional forms for the drift and diffusion 
coefficients from their low-frequency forms. 

Next, we developed theoretical expressions for the mean torque and diffusion
coefficients for several external processes, including collisions of gas-phase
particles with the grain, thermal evaporation from the grain surface, and 
the formation of H$_2$ molecules, followed by their ejection from the grain
surface. These apply for the special case of a spheroidal grain with a
non-uniform mass density (with the center of mass at the center of the 
spheroid and the principal axis of greatest moment of inertia lying along
the spheroid symmetry axis). We adopted several simplifications in the 
analysis of H$_2$-formation. The translational kinetic energy of the ejected
molecules is taken to be constant. In the case of special formation sites on
the grain surface, the ejection rate is taken to be equal at all of the sites
and the molecules depart along a single direction at each site. 

From large simulation suites, in which the Langenvin equations for both
internal and external processes are integrated, we reach the following 
conclusions. First, the mean duration of up-steps (when the systematic torque 
acts so as to spin the grain up) exceeds the mean duration of down-steps.
Second, thermal trapping is not prevalent during crossovers. Third, the
Davis-Greenstein mechanism, with suprathermal spin-up, can drive grains into
alignment in the cold neutral medium, without significant impediment from 
thermal trapping. However, it does not appear to be a viable explanation of
grain alignment in the diffuse ISM, at least for the relatively large 
grains that are responsible for optical and infrared starlight polarization, 
since the alignment time-scale is long. The D-G mechanism could, however,
potentially yield some small-grain alignment, with observable consequences
for ultraviolet starlight polarization. Future work will examine this 
possibility in greater detail.

Currently, the consensus view is that radiative torques dominate in the 
alignment of relatively large grains; see \citet{L07} and \citet{ALV15}
for reviews. This view has resulted, in part, from the conclusion of 
\citet{LD99b} that grains subjected only to torques that are fixed relative 
the grain body are thermally trapped. We conclude that, even without thermal
trapping, D-G alignment does not effectively align large grains. 

Detailed models have found that, 
in the radiative-torque alignment scenario, grains can pass through 
crossovers and can reach aligned states characterized by either suprathermal
or thermal rotation \citep{WD03, HL09}. Thus, the main result of this paper,
that the mean duration of up-steps exceeds the mean duration of down steps,
could have significant implications for radiative-torque alignment as well 
as for D-G alignment. In future work, we will adapt the computational and
theoretical tools developed here to a study of radiative-torque alignment.
We will also consider grains with superparamagnetic inclusions. 

Finally, \citet{Purcell79} focused on grain alignment, but also noted that
the tensile stress within a suprathermally rotating grain could possibly 
disrupt the grain, depending on its structure. Recently, Hoang and 
collaborators \citep[e.g.][]{HTLA19, H19, H20, LH20} have developed this idea
in detail, with a focus on radiative torques. The grain equilibrium rotational 
rates for the model of H$_2$-formation torques adopted here are shown in 
Fig.~\ref{fig:Jp-eq-hist} and equation (\ref{eq:Jp_eq_from_Qs}). Comparing
with Fig.~13 in \citet{LH20} and Figs.~12--15 in \citet{DW97}, these are
comparable to results from previous studies of H$_2$-formation torques and 
can exceed the rotational rates arising from radiative torques, for conditions
typical of the CNM. Thus, our conclusion that thermal trapping is not prevalent 
could have significant implications for the rotational disruption of grains,
especially in environments where radiative torques are weak. See Section 9.4
in \citet{LH20} for more details. The uncertainties associated with the
H$_2$-formation model, noted at the end of Section \ref{sec:D-G-alignment}
above, will need to be resolved in order to clarify the importance of the
resulting torques to grain disruption. 

\section*{Acknowledgements}

We are grateful to Bruce Draine and the anonymous referee for helpful comments
on the manuscript.

\section*{Data Availability}

The data underlying this article will be shared on reasonable request to the
corresponding author.

\bibliographystyle{mnras}
\bibliography{mybib}

\begin{thebibliography}{}
\makeatletter
\relax
\def\mn@urlcharsother{\let\do\@makeother \do\$\do\&\do\#\do\^\do\_\do\%\do\~}
\def\mn@doi{\begingroup\mn@urlcharsother \@ifnextchar [ {\mn@doi@}
  {\mn@doi@[]}}
\def\mn@doi@[#1]#2{\def\@tempa{#1}\ifx\@tempa\@empty \href
  {http://dx.doi.org/#2} {doi:#2}\else \href {http://dx.doi.org/#2} {#1}\fi
  \endgroup}
\def\mn@eprint#1#2{\mn@eprint@#1:#2::\@nil}
\def\mn@eprint@arXiv#1{\href {http://arxiv.org/abs/#1} {{\tt arXiv:#1}}}
\def\mn@eprint@dblp#1{\href {http://dblp.uni-trier.de/rec/bibtex/#1.xml}
  {dblp:#1}}
\def\mn@eprint@#1:#2:#3:#4\@nil{\def\@tempa {#1}\def\@tempb {#2}\def\@tempc
  {#3}\ifx \@tempc \@empty \let \@tempc \@tempb \let \@tempb \@tempa \fi \ifx
  \@tempb \@empty \def\@tempb {arXiv}\fi \@ifundefined
  {mn@eprint@\@tempb}{\@tempb:\@tempc}{\expandafter \expandafter \csname
  mn@eprint@\@tempb\endcsname \expandafter{\@tempc}}}

\bibitem[\protect\citeauthoryear{{Andersson}, {Lazarian}  \&
  {Vaillancourt}}{{Andersson} et~al.}{2015}]{ALV15}
{Andersson} B.~G.,  {Lazarian} A.,   {Vaillancourt} J.~E.,  2015, \mn@doi
  [\araa] {10.1146/annurev-astro-082214-122414}, \href
  {https://ui.adsabs.harvard.edu/abs/2015ARA&A..53..501A} {53, 501}

\bibitem[\protect\citeauthoryear{{Davis} \& {Greenstein}}{{Davis} \&
  {Greenstein}}{1951}]{DG51}
{Davis} Jr. L.,  {Greenstein} J.~L.,  1951, \mn@doi [\apj] {10.1086/145464},
  \href {http://adsabs.harvard.edu/abs/1951ApJ...114..206D} {114, 206}

\bibitem[\protect\citeauthoryear{{Draine} \& {Weingartner}}{{Draine} \&
  {Weingartner}}{1997}]{DW97}
{Draine} B.~T.,  {Weingartner} J.~C.,  1997, \apj, \href
  {http://adsabs.harvard.edu/abs/1997ApJ...480..633D} {480, 633}

\bibitem[\protect\citeauthoryear{{Hoang}}{{Hoang}}{2019}]{H19}
{Hoang} T.,  2019, \mn@doi [\apj] {10.3847/1538-4357/ab1075}, \href
  {https://ui.adsabs.harvard.edu/abs/2019ApJ...876...13H} {876, 13}

\bibitem[\protect\citeauthoryear{{Hoang}}{{Hoang}}{2020}]{H20}
{Hoang} T.,  2020, \mn@doi [Galaxies] {10.3390/galaxies8030052}, \href
  {https://ui.adsabs.harvard.edu/abs/2020Galax...8...52H} {8, 52}

\bibitem[\protect\citeauthoryear{{Hoang} \& {Lazarian}}{{Hoang} \&
  {Lazarian}}{2009}]{HL09}
{Hoang} T.,  {Lazarian} A.,  2009, \mn@doi [\apj]
  {10.1088/0004-637X/695/2/1457}, \href
  {http://adsabs.harvard.edu/abs/2009ApJ...695.1457H} {695, 1457}

\bibitem[\protect\citeauthoryear{{Hoang}, {Lazarian}  \& {Martin}}{{Hoang}
  et~al.}{2014}]{HLM14}
{Hoang} T.,  {Lazarian} A.,   {Martin} P.~G.,  2014, \mn@doi [\apj]
  {10.1088/0004-637X/790/1/6}, \href
  {https://ui.adsabs.harvard.edu/abs/2014ApJ...790....6H} {790, 6}

\bibitem[\protect\citeauthoryear{{Hoang}, {Tram}, {Lee}  \& {Ahn}}{{Hoang}
  et~al.}{2019}]{HTLA19}
{Hoang} T.,  {Tram} L.~N.,  {Lee} H.,   {Ahn} S.-H.,  2019, \mn@doi [Nature
  Astronomy] {10.1038/s41550-019-0763-6}, \href
  {https://ui.adsabs.harvard.edu/abs/2019NatAs...3..766H} {3, 766}

\bibitem[\protect\citeauthoryear{{Jones} \& {Spitzer}}{{Jones} \&
  {Spitzer}}{1967}]{JS67}
{Jones} R.~V.,  {Spitzer} Lyman J.,  1967, \mn@doi [\apj] {10.1086/149086},
  \href {https://ui.adsabs.harvard.edu/abs/1967ApJ...147..943J} {147, 943}

\bibitem[\protect\citeauthoryear{{Kim} \& {Martin}}{{Kim} \&
  {Martin}}{1995}]{KM95}
{Kim} S.-H.,  {Martin} P.~G.,  1995, \mn@doi [\apj] {10.1086/175604}, \href
  {https://ui.adsabs.harvard.edu/abs/1995ApJ...444..293K} {444, 293}

\bibitem[\protect\citeauthoryear{{Kolasi} \& {Weingartner}}{{Kolasi} \&
  {Weingartner}}{2017}]{KW17}
{Kolasi} E.,  {Weingartner} J.~C.,  2017, \mn@doi [\mnras]
  {10.1093/mnras/stx1694}, \href
  {http://adsabs.harvard.edu/abs/2017MNRAS.471.1222K} {471, 1222}

\bibitem[\protect\citeauthoryear{{Lazarian}}{{Lazarian}}{2007}]{L07}
{Lazarian} A.,  2007, \mn@doi [\jqsrt] {10.1016/j.jqsrt.2007.01.038}, \href
  {https://ui.adsabs.harvard.edu/abs/2007JQSRT.106..225L} {106, 225}

\bibitem[\protect\citeauthoryear{{Lazarian} \& {Draine}}{{Lazarian} \&
  {Draine}}{1997}]{LD97}
{Lazarian} A.,  {Draine} B.~T.,  1997, \mn@doi [\apj] {10.1086/304587}, \href
  {http://adsabs.harvard.edu/abs/1997ApJ...487..248L} {487, 248}

\bibitem[\protect\citeauthoryear{{Lazarian} \& {Draine}}{{Lazarian} \&
  {Draine}}{1999a}]{LD99a}
{Lazarian} A.,  {Draine} B.~T.,  1999a, \mn@doi [\apjl] {10.1086/311986}, \href
  {http://adsabs.harvard.edu/abs/1999ApJ...516L..37L} {516, L37}

\bibitem[\protect\citeauthoryear{{Lazarian} \& {Draine}}{{Lazarian} \&
  {Draine}}{1999b}]{LD99b}
{Lazarian} A.,  {Draine} B.~T.,  1999b, \mn@doi [\apjl] {10.1086/312137}, \href
  {http://adsabs.harvard.edu/abs/1999ApJ...520L..67L} {520, L67}

\bibitem[\protect\citeauthoryear{Lazarian \& Hoang}{Lazarian \&
  Hoang}{2020}]{LH20}
Lazarian A.,  Hoang T.,  2020, Alignment and rotational disruption of dust
  (\mn@eprint {arXiv} {2010.15301})

\bibitem[\protect\citeauthoryear{{Lee} \& {Draine}}{{Lee} \&
  {Draine}}{1985}]{LD85}
{Lee} H.~M.,  {Draine} B.~T.,  1985, \mn@doi [\apj] {10.1086/162974}, \href
  {https://ui.adsabs.harvard.edu/abs/1985ApJ...290..211L} {290, 211}

\bibitem[\protect\citeauthoryear{{Mathis}}{{Mathis}}{1986}]{M86}
{Mathis} J.~S.,  1986, \mn@doi [\apj] {10.1086/164499}, \href
  {https://ui.adsabs.harvard.edu/abs/1986ApJ...308..281M} {308, 281}

\bibitem[\protect\citeauthoryear{{Peters} et~al.,}{{Peters}
  et~al.}{2017}]{Peters17}
{Peters} T.,  et~al., 2017, \mn@doi [\mnras] {10.1093/mnras/stx341}, \href
  {https://ui.adsabs.harvard.edu/abs/2017MNRAS.467.4322P} {467, 4322}

\bibitem[\protect\citeauthoryear{Press, Teukolsky, Vetterling  \&
  Flannery}{Press et~al.}{1992}]{Press92}
Press W.~H.,  Teukolsky S.~A.,  Vetterling W.~T.,   Flannery B.~P.,  1992,
  Numerical Recipes in FORTRAN; The Art of Scientific Computing, 2nd edn.
Cambridge University Press, New York, NY, USA

\bibitem[\protect\citeauthoryear{{Purcell}}{{Purcell}}{1979}]{Purcell79}
{Purcell} E.~M.,  1979, \mn@doi [\apj] {10.1086/157204}, \href
  {http://adsabs.harvard.edu/abs/1979ApJ...231..404P} {231, 404}

\bibitem[\protect\citeauthoryear{{Purcell} \& {Spitzer}}{{Purcell} \&
  {Spitzer}}{1971}]{PS71}
{Purcell} E.~M.,  {Spitzer} Jr. L.,  1971, \mn@doi [\apj] {10.1086/151002},
  \href {http://adsabs.harvard.edu/abs/1971ApJ...167...31P} {167, 31}

\bibitem[\protect\citeauthoryear{{Roberge} \& {Lazarian}}{{Roberge} \&
  {Lazarian}}{1999}]{RL99}
{Roberge} W.~G.,  {Lazarian} A.,  1999, \mn@doi [\mnras]
  {10.1046/j.1365-8711.1999.02464.x}, \href
  {http://adsabs.harvard.edu/abs/1999MNRAS.305..615R} {305, 615}

\bibitem[\protect\citeauthoryear{{Roberge}, {Degraff}  \& {Flaherty}}{{Roberge}
  et~al.}{1993}]{RDF93}
{Roberge} W.~G.,  {Degraff} T.~A.,   {Flaherty} J.~E.,  1993, \mn@doi [\apj]
  {10.1086/173390}, \href {http://adsabs.harvard.edu/abs/1993ApJ...418..287R}
  {418, 287}

\bibitem[\protect\citeauthoryear{{Spitzer} \& {McGlynn}}{{Spitzer} \&
  {McGlynn}}{1979}]{SM79}
{Spitzer} Jr. L.,  {McGlynn} T.~A.,  1979, \mn@doi [\apj] {10.1086/157205},
  \href {http://adsabs.harvard.edu/abs/1979ApJ...231..417S} {231, 417}

\bibitem[\protect\citeauthoryear{{Weingartner}}{{Weingartner}}{2009}]{W09}
{Weingartner} J.~C.,  2009, \mn@doi [\apj] {10.1088/0004-637X/690/1/875}, \href
  {http://adsabs.harvard.edu/abs/2009ApJ...690..875W} {690, 875}

\bibitem[\protect\citeauthoryear{{Weingartner} \& {Draine}}{{Weingartner} \&
  {Draine}}{2001}]{WD01force}
{Weingartner} J.~C.,  {Draine} B.~T.,  2001, \mn@doi [\apj] {10.1086/320963},
  \href {https://ui.adsabs.harvard.edu/abs/2001ApJ...553..581W} {553, 581}

\bibitem[\protect\citeauthoryear{{Weingartner} \& {Draine}}{{Weingartner} \&
  {Draine}}{2003}]{WD03}
{Weingartner} J.~C.,  {Draine} B.~T.,  2003, \mn@doi [\apj] {10.1086/374597},
  \href {http://adsabs.harvard.edu/abs/2003ApJ...589..289W} {589, 289}

\makeatother
\end{thebibliography}

\bsp	

\label{lastpage}
\end{document}